\documentclass[manuscript]{emulateapj}
\usepackage{graphics}
\usepackage{graphicx}
\usepackage{rotating}
\usepackage{mathtools}
\usepackage{footmisc}
\usepackage{xhfill}
\usepackage[T1]{fontenc}
\begin{document}

\newcommand{\kms}{\mbox{km~s$^{-1}$}}
\newcommand{\s}{\mbox{$''$}}
\newcommand{\mloss}{\mbox{$\dot{M}$}}
\newcommand{\mdot}{\mbox{$\dot{M}$}}
\newcommand{\my}{\mbox{$M_{\odot}$~yr$^{-1}$}}
\newcommand{\ls}{\mbox{$L_{\odot}$}}
\newcommand{\um}{\mbox{$\mu$m}}
\newcommand{\ujy}{\mbox{$\mu$Jy}}
\newcommand{\ms}{\mbox{$M_{\odot}$}}
\newcommand{\rsun}{\mbox{$R_{\odot}$}}

\newcommand{\vexp}{\mbox{$V_{\rm exp}$}}
\newcommand{\vsys}{\mbox{$V_{\rm sys}$}}
\newcommand{\vlsr}{\mbox{$V_{\rm LSR}$}}
\newcommand{\tex}{\mbox{$T_{\rm ex}$}}
\newcommand{\teff}{\mbox{$T_{\rm eff}$}}
\newcommand{\tmb}{\mbox{$T_{\rm mb}$}}
\newcommand{\trot}{\mbox{$T_{\rm rot}$}}
\newcommand{\tkin}{\mbox{$T_{\rm kin}$}}
\newcommand{\dens}{\mbox{$n_{\rm H_2}$}}
\newcommand{\bri}{\mbox{erg\,s$^{-1}$\,cm$^{-2}$\,\AA$^{-1}$\,arcsec$^{-2}$}}
\newcommand{\brib}{\mbox{erg\,s$^{-1}$\,cm$^{-2}$\,arcsec$^{-2}$}}
\newcommand{\flux}{\mbox{erg\,s$^{-1}$\,cm$^{-2}$\,\AA$^{-1}$}}
\newcommand{\ha}{\mbox{H$\alpha$}}
\newcommand{\codos}{$^{12}$CO}
\newcommand{\cotres}{$^{13}$CO}
\newcommand{\hcnmol}{H$^{13}$CN}
\newcommand{\paani}{H$_2$O}
\newcommand{\soocho}{SO\,N,J=8,8-7,7}
\newcommand{\csocho}{CS\,J=7-6}
\newcommand{\siodonau}{$^{29}$SiO\,J=8-7\,(v=0)}
\newcommand{\sodos}{SO$_2$\,4(3,1)-3(2,2)}
\newcommand{\soteenchar}{$^{34}$SO\,N,J=8,7-7,6}
\newcommand{\pscalunit}{g\,cm\,s$^{-1}$}
\newcommand{\cdensunit}{cm$^{-2}$}
\newcommand{\vdensunit}{cm$^{-3}$}
\newcommand{\engyunit}{erg}
\newcommand{\hmol}{H$_2$}
\newcommand{\intunit}{erg\,s$^{-1}$\,cm$^{-2}$\,\AA$^{-1}$}
\newcommand{\jykmsperbeam}{Jy-km\,s$^{-1}$\,beam$^{-1}$}

\newcommand{\ditto}[1][.3pt]{\xrfill{#1}~\textquotedbl~\xrfill{#1}}
\newcommand{\dittodos}[1][.3pt]{\textquotedbl}

\title{The Coldest Place in the Universe: Probing the Ultra-Cold Outflow and Dusty
Disk in the Boomerang Nebula}
\author{R. Sahai\altaffilmark{1}, W.H.T. Vlemmings\altaffilmark{2},  L-\AA. Nyman\altaffilmark{3}
}

\altaffiltext{1}{Jet Propulsion Laboratory, MS\,183-900, California
Institute of Technology, Pasadena, CA 91109, USA}
\altaffiltext{2}{Department of Earth and Space Sciences, Chalmers University of Technology, Onsala Space Observatory, SE-43992 Onsala, Sweden}
\altaffiltext{3}{Joint ALMA Observatory (JAO), Alonso de Cordova 3107, Vitacura, Chile, and 
European Southern Observatory, Alonso de Cordova 3107, Vitacura, Santiago, Chile}

\email{raghvendra.sahai@jpl.nasa.gov}
\begin{abstract}
Our Cycle 0 ALMA observations confirmed that the Boomerang Nebula is the coldest known object in the Universe, with a massive high-speed outflow that 
has cooled significantly below the cosmic background temperature. Our new CO 1--0 data reveal heretofore unseen distant regions of this ultra-cold 
outflow, out to $\gtrsim120,000$\,AU. We find that in the ultra-cold outflow, the mass-loss rate ($\mdot$) increases with radius, similar 
to its expansion velocity ($V$) -- taking $V\propto r$, we find $\mdot \propto r^{0.9-2.2}$. The mass in the ultra-cold outflow is 
$\gtrsim3.3$\,\ms, and the Boomerang's main-sequence progenitor mass is $\gtrsim4$\,\ms. Our high angular resolution ($\sim0\farcs3$) CO 
J=3--2 map shows the inner bipolar nebula's precise, highly-collimated shape, and a dense central waist of size (FWHM) 
$\sim$1740\,AU$\times275$\,AU. The molecular gas and the dust as seen in scattered light via optical HST 
imaging show a detailed correspondence. The waist shows a compact core in thermal dust emission at 0.87--3.3\,mm, which harbors  
$(4-7)\times10^{-4}$\,\ms~of very large ($\sim$mm-to-cm sized), cold ($\sim20-30$\,K) grains. The central waist (assuming its outer regions to be 
expanding) and fast bipolar outflow have expansion ages of $\lesssim1925$\,yr and $\le1050$\,yr: the ``jet-lag'' (i.e., torus age minus the 
fast-outflow age) in the Boomerang supports models in which the primary star interacts directly with a binary companion. We argue that this 
interaction resulted in a common-envelope configuration while the Boomerang's primary was an RGB or early-AGB star, with the companion finally merging 
into the primary's core, and ejecting the primary's envelope that now forms the ultra-cold outflow.
\end{abstract}


\section{Introduction}
The Boomerang Nebula (Wegner \& Glass 1979), is a bipolar pre-planetary Nebula (PPN). PPNe are generally believed to represent a short-lived ($\sim 
1000$ yrs) transition
phase during which Asymptotic Giant Branch (AGB) stars and their round circumstellar envelopes (CSEs) evolve into pre-planetary and planetary nebulae 
(PNe) with a breathtaking variety of aspherical geometrical shapes and symmetries (e.g., Sahai et al. 2007, Sahai, Morris \& Villar 2011). The 
Boomerang 
holds the distinction of being the (naturally occurring) coldest known object in the Universe 
(Sahai \& Nyman 1997: SN97). The Boomerang's very large mass-loss rate ($\sim0.001$\,\my) and low-luminosity (300\ls) are 
unprecedented, making it a key object for testing theoretical models (i) for mass-loss during post-main sequence evolution 
(e.g., Woitke 2006, Winters et al. 2000), and (ii) for producing
the dazzling variety of bipolar and multipolar morphologies seen in PNe (Balick \& Frank 2002).

Single dish CO J=1--0 observations showed an extended high-speed outflow in absorption against the microwave background, 
implying that the nebula has cooled to a temperature
significantly below that of the cosmic background radiation ($T_{bb}=2.7$\,K) due to adiabatic expansion, 
but the structure of the outflow was not properly resolved due to the large beam-size (45\arcsec). SN97 modeled the objects as consisting of 
two nested spherically symmetric shells: a warm inner shell extending 2.5\arcsec--6\arcsec~with an expansion
velocity of about 35\,\kms, and an ultra-cold, extended outer shell extending 6\arcsec--33\arcsec, with a
velocity of about 164\,\kms. 

We mapped the Boomerang with ALMA in Cycle 0 
in the CO J=1--0 and 2--1 lines using the compact configuration (i.e., with 4.3\arcsec\, and 2.2\arcsec\ resolution)
to determine the morphologies of its molecular outflows and compare it with the optical images (Sahai et al. 2013: Setal13). We 
confirmed that the extended high-velocity envelope has ultra-cold temperatures below the CMB, and 
although patchy, shows no systematic departures from a roughly round shape, and the observed angular sizes of the inner and outer outflows were in 
reasonable agreement with the SN97 model. The inner outflow was found to have 
an hourglass shape with a dense central waist. The waist shapes the illumination of 
the nebula as a whole and provides a natural explanation for the overall hourglass optical morphology.
We discovered patchy emission around the ultra-cold outflow, as expected from eventual heating of the gas due to grain photoelectric heating. 
Continuum emission at 1.3 and 2.6\,mm indicated the presence of a substantial mass of mm-sized grains in the central dense waist.

Here, we report the highest-angular molecular-line observations of this object with ALMA at 0.88\,mm,  
new CO 1--0 data that recover all of the flux from the very extended ultra-cold outflow, and continuum observations spanning 
the 0.88--3.3\,mm wavelength range that robustly constrain the propertes of the central dust source. 
A preliminary version of some of the observational results from these data was given in Sahai et al. (2015).

The plan of the paper is as follows. In \S\,\ref{obs} we describe the observational setups, and data reduction and calibration procedures. In 
\S\,\ref{result} we present our main observational results. These include: the molecular-gas morphology and kinematics of the central bipolar nebula and its dense waist, 
and a comparison with archival HST imaging in polarized light (\S\,\ref{coj32}), the physical properties of the ultra-cold outflow (\S\,\ref{coj10}), 
the unexpected discovery of SO emission (\S\,\ref{findso}), 
the presence of very large grains in the central waist (\S\,\ref{cont_txt}), and the nebular polarization properties (\S\,\ref{polr}). In \S\,\ref{discus}, we discuss the 
implications of important time-scales in the Boomerang derived from 
our data for theoretical models for the formation of jets and torii in AGB and post-AGB objects. We argue that the extreme properties of the 
ultra-cold outflow imply that a common-envelope event produced the Boomerang Nebula
The main conclusions of our study are summarized in \S\,\ref{conclude}. 

Although we have adopted a distance of D=1.5\,kpc for the Boomerang (as inferred by SN97) for most of our analysis in this paper, we discuss 
the distance uncertainty where relevant.

\section{Observations}\label{obs}
A summary of the continuum and spectral line observations of the Boomerang Nebula is given in Table\,\ref{obslog}. The continuum 
Band 3 and CO J=3--2 Band 7 observations were taken using a single pointing centered on R.A.$=12^h44^m45.978^s$ and 
$\delta=-54^\circ31'13.043{''}$. The \codos~J=1--0 Band 3 observations consisted of a 7-point mosaic with the ALMA 12-m array 
and a 5-point mosaic using the Atacama Compact Array (ACA) centered on the same position. Additionally on-the-fly (OTF) 
mapping observations using three ALMA Total Power (TP) antennas, using a total of 160~min, was taken. The 12-m array 
observations were completed on December 17 2013, the ACA observations on March 20-21 2014 and the TP observations on 
April 25-26 2015. By combining the 12-m, ACA and TP data, the observations are sensitive to emission on all scales, with 
baselines ranging up to 1280~m. Flux calibration was done using Callisto (12-m Band 3 spectral line observations), Mars 
(ACA and 12-m continuum Band 3 observations), and Titan (12-m Band 7 observations). Bandpass and gain calibrations were 
performed using the quasars J1037-2934 and J1112-5703 for the 12-m Band 3 spectral line observations, and J1107-449 and 
J1303-5540 for all other observations. The data were calibrated using the Common Astronomy Software Application (CASA 
4.2) and the observatory provided scripts with only a few additional flagging commands. The Band 3 data of the 12-m and 
ACA observations were combined in the visibility plane and imaged using natural weighting and a Gaussian taper radius of 
100~$k\lambda$. Subsequently, the TP image, also obtained using the provided script, was combined with the 
interferometric image using the CASA task {\it feather}. The resulting beam sizes and position angles are provided in Table\,\ref{obslog}. 

In order to assess possible flux loss in the interferometric \codos~J=3-2 observations, we obtained a spectrum of the line 
on October 8 2014 using the APEX telescope. The observations were done in beam-switching mode using the SHeFI-APEX2 
receiver. 

\section{Results}\label{result}

\subsection{CO\,(J=3-2) and Optical Imaging}\label{coj32}

\subsubsection{Spatio-Kinematic Structure of the Lobes}
\begin{figure}[!ht]
\includegraphics[trim=0.0in 1.1in 1.8in 0.5in,clip,width=0.8\textwidth]{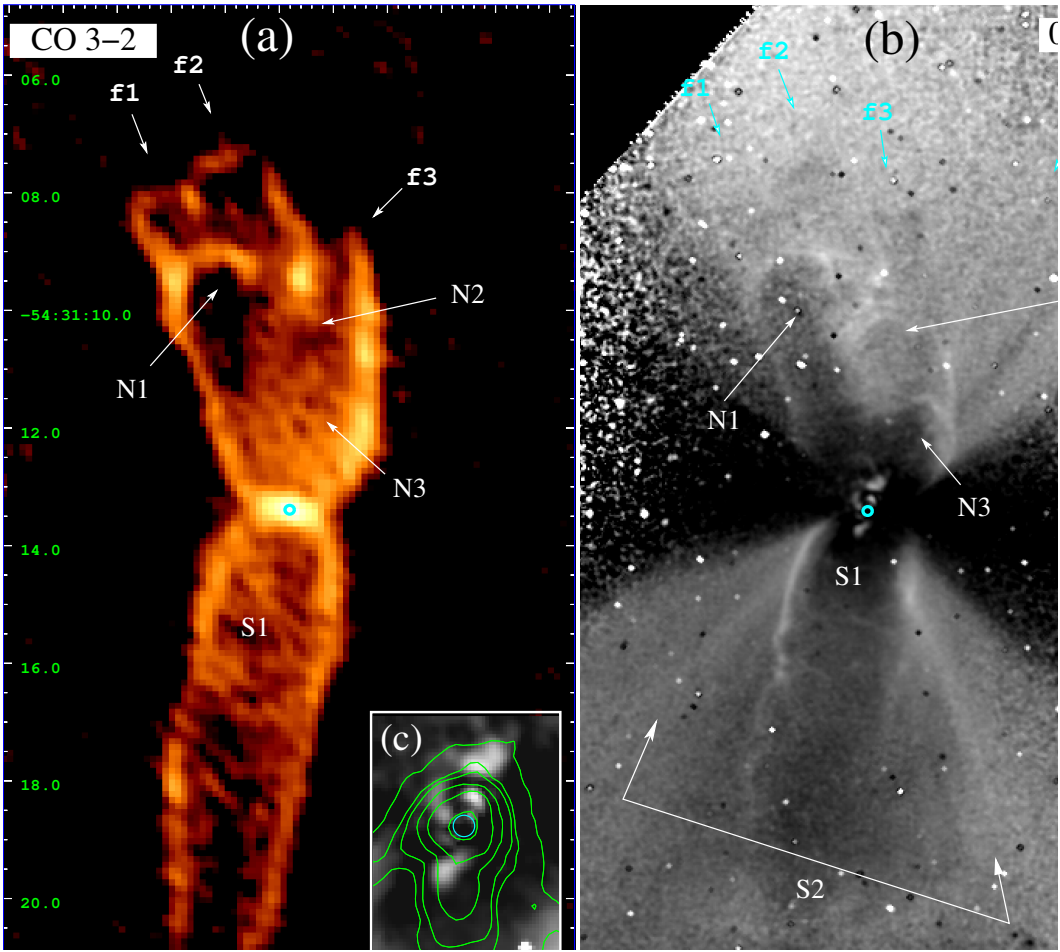}
\caption{The central bipolar nebula in the Boomerang (a) ALMA $^{12}$CO J=3--2 moment zero map (integrated over the velocity
range of $-21.5<V_{lsr}\,\kms<6.5$, and (b) HST image of the 0.6\micron~percentage polarization. Inset (c) 
shows expanded view of the central region $0\farcs65\times0\farcs95$ region, with contours showing the total 0.6\micron~intensity at 
0.47, 0.87, 1.1, 3.3, and 32\%, of the peak. Cyan circles in the panels mark the location of the 0.87\micron~continuum peak, 
at (J2000) RA=12:44:46.084, Dec=-54:31:13.35.}
\vspace{-0.1in}
\label{boom-acs-co32}
\end{figure}

The J=3--2 line of both $^{12}$CO and $^{13}$CO were observed with the 12m array with an angular resolution of about 0\farcs3 (Table\,\ref{obslog}). 
Our CO J=3--2 integrated-intensity image ($-21.5<V_{lsr}\,(kms)<7.0$) shows the precise, highly collimated shape of an inner bipolar structure and a 
dense central waist, with unprecedented angular resolution ($0.4{''}$ or 600\,AU) (Fig.\,\ref{boom-acs-co32}a). 

For the first time, we have adequate angular resolution at millimeter-wavelengths to enable detailed matching between the 
molecular-line structure of the lobes in a PPN to the structure seen in scattered light in the optical. A comparison of the CO J=3--2 image 
with a 0.6\,\micron~polarized-light image obtained using the ACS  
(Fig.\,\ref{boom-acs-co32}b, Cracraft \& Sparks 2007), 
confirms the common interpretation of these as expanding, thin-walled cavities. Since the absolute astrometry of the HST image is not 
sufficiently accurate, we have registered it relative to the ALMA images by forcing 
the 0.6\,\micron~continuum peak to coincide with the 0.87\micron~continuum peak.

Both the CO J=3--2 and the optical images shows that the N-lobe has a much 
more complex geometrical structure than the S-lobe. In the optical percentage polarized image, 
at least 3 limb-brightened, bubble-like structures can be clearly seen in the N-lobe (N1, N2, 
N3), and additional, much fainter structures can be seen beyond these (f1, f2, f3, and f4). In contrast, the S-lobe shows a single, 
bright cylindrical-shaped structure (S1), within a conical structure (S2). The N1, N2, and N3 bubbles merge together 
in one large limb-brightened structure in the CO J=3--2 image. The f1 and f2 bubbles are seen much more clearly in the 
CO image than in the optical image. A more detailed discussion of the polarization in the Boomerang is provided in S\,\ref{polr}.

\begin{figure}[!ht]
\resizebox{1.0\textwidth}{!}{\includegraphics{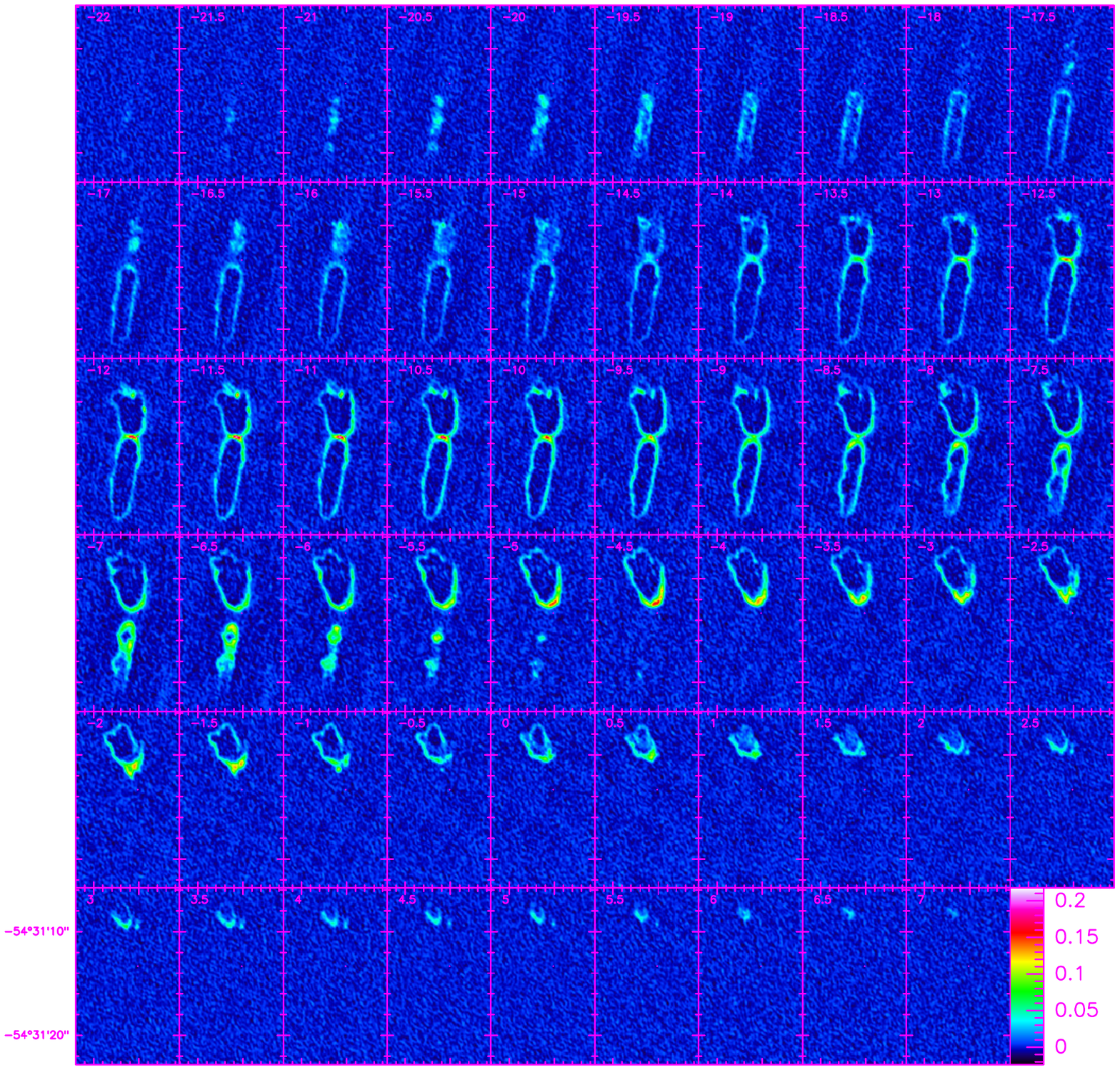}}
\caption{CO J=3--2 channel map of the Boomerang. Intensity scale-bar units are Jy beam$^{-1}$. The 1$\sigma$ noise is 
$4.8\times10^{-3}$ Jy beam$^{-1}$.}
\label{12co32panel}
\end{figure}

\begin{figure}[!ht]
\resizebox{1.0\textwidth}{!}{\includegraphics{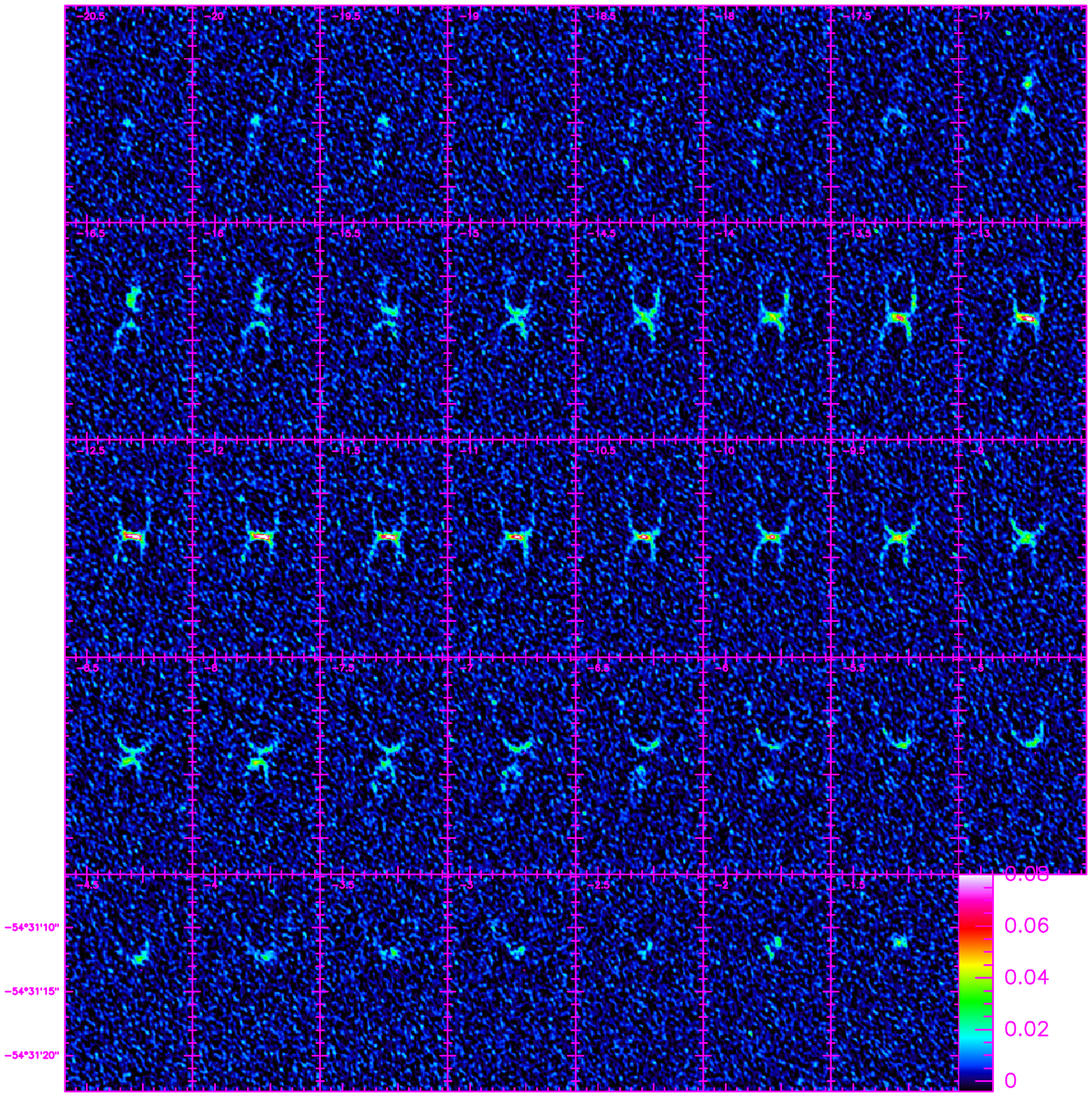}}
\caption{$^{13}$CO J=3--2 channel map of the Boomerang. Intensity scale-bar units are Jy beam$^{-1}$. The 1$\sigma$ 
noise is $5.3\times10^{-3}$ Jy beam$^{-1}$}
\label{13co32panel}
\end{figure}

The channel maps of the J=3--2 $^{12}$CO and $^{13}$CO emission are shown in (Figs.\,\ref{12co32panel},\,\ref{13co32panel}), and reveal the
3-dimensional structure of the central bipolar nebula that was inferred from the lower angular-resolution J=2--1 $^{12}$CO map by Setal13. The 
$^{13}$CO map shows a similar structure as $^{12}$CO, but has a significantly lower S/N ratio, only the base and mid-latitude regions of the lobes 
are detected. The channels near the 
systemic velocity, $V_{lsr}\sim-11.5$\,\kms~represent a cross-sectional cut through the lobes perpendicular to the line-of-sight ($los$). The spatial narrowing of 
the emission in the southern lobe (S-lobe) towards its axis as one approaches the blue- and red-shifted channels ($V_{lsr}\sim-20$\,\kms~and 
$V_{lsr}\sim-6$\,\kms) is expected, as these channels sample material in the lobe walls that is moving directly towards or away from us.

As noted by Setal13, the lobes are not symmetrical -- first, the S-lobe is much more elongated than the northern one (N-lobe), and second the  
symmetry axes of these lobes are not colinear. The N-lobe has a significantly more complex spatio-kinematic structure than the S-lobe, 
as also seen in the HST image. The N-lobe's axis is well-aligned with that of the S-lobe (at $PA\sim-8\arcdeg$) in 
the blue-shifted channels in the $V_{lsr}\sim-17.5$ to $-14.5$\,\kms~range, and the progressive spatial narrowing of the emission towards the axis 
is similar to that seen for the sourthern lobe. However, in channels redwards of $V_{lsr}\sim-14.5$\,\kms, the N-lobe's axis tilts 
progressively towards large PA values ($29\arcdeg$), and is dominated by the change in orientation of its eastern wall. The western wall of the 
N-lobe is less affected in this manner, especially its low-latitude basal region towards the west, which corresponds to the western wall of the N3 
optical lobe. 

We derive an upper limit of 1050\,yr to the expansion time-scale of the S-lobe by dividing its linear extent, 7\farcs9 
or $1.8\times10^{17}$\,cm, by 
the deprojected expansion velocity, $\ge14$\,\kms/sin($\theta$), where $90\arcdeg>\theta\ge75\arcdeg$ is the lobe's inclination\footnote{all inclination 
angles, 
here and elsewhere in the paper, are relative to the $los$} (Setal15). The expansion 
time-scale of the N-lobe is comparable or smaller, as its linear extent is somewhat smaller, but since its structure is significantly 
more complex, we cannot constrain it as well as for the S-lobe.

We obtained a single-dish spectrum of the CO J=3--2 emission from the Boomerang with the 17\farcs8 beam of the 
12-m APEX (Atacama Pathfinder EXperiment) telescope (Fig.\,\ref{co32apex}). 
The peak intensity is about 0.2 K, implying a total flux of about 8.2 Jy. The ALMA observations thus resolve out about 4 Jy of extended emission. 



\subsubsection{Spatio-Kinematic Structure of the Waist}
\begin{figure}[!ht]
\resizebox{1.0\textwidth}{!}{\includegraphics{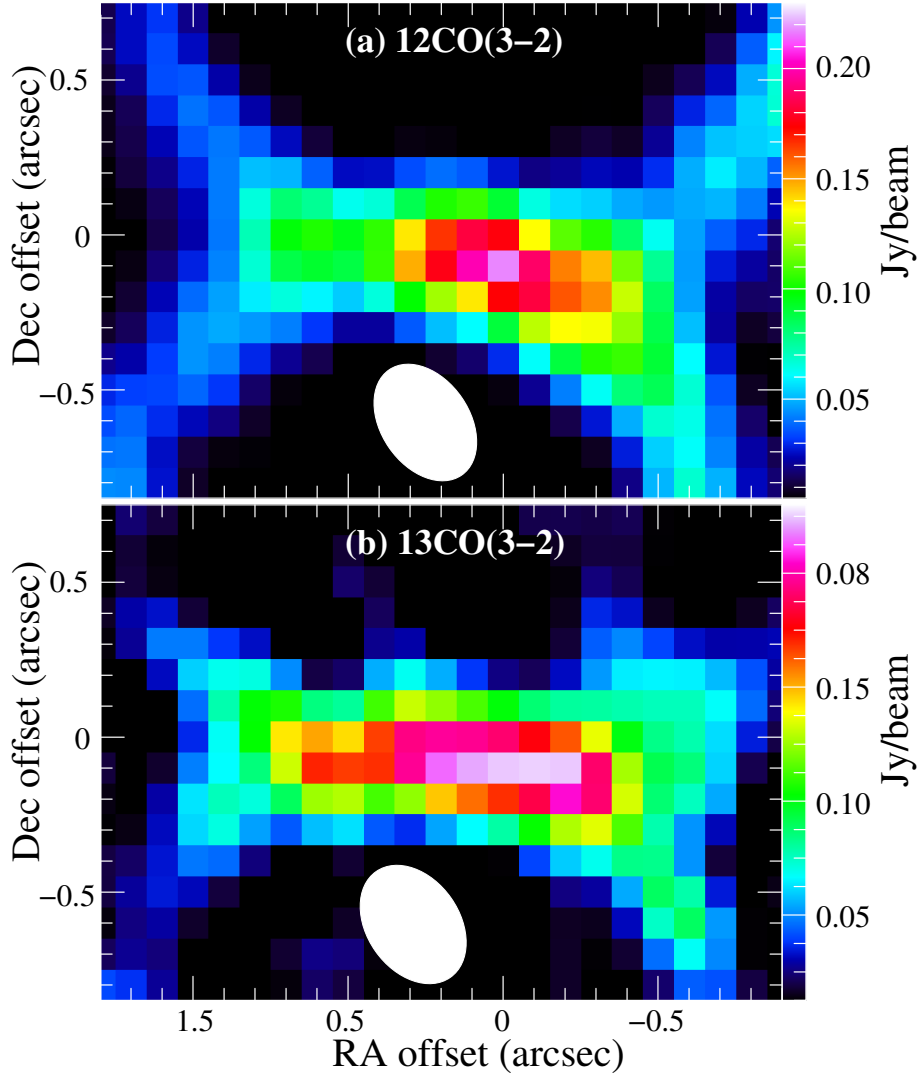}}
\caption{The waist of the Boomerang Nebula, as seen in (a) $^{12}$CO J=3--2 
(white ellipse: beam FWHM $0\farcs37\times0\farcs25$, $PA=34.4$\arcdeg), and (b) $^{13}$CO J=3--2 
(white ellipse: beam FWHM $0\farcs37\times0\farcs26$, $PA=34.4$\arcdeg) images, at the central velocity of the waist emission profile, $V_{lsr}=-12$\,kms.}
\label{co32waist}
\end{figure}

\begin{figure}[!ht]
\rotatebox{270}{\resizebox{0.5\textwidth}{!}{\includegraphics{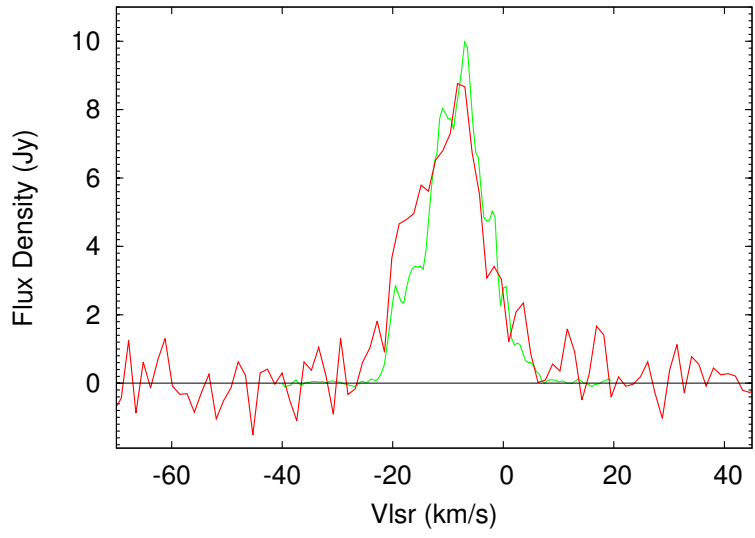}}}
\vskip 0.5in
\caption{Single-dish (APEX) CO J=3--2 spectrum of the Boomerang, compared to the spatially-integrated ALMA spectrum (scaled up by factor 2).}
\label{co32apex}
\end{figure}

The central waist, which is seen nearly edge-on in the $^{12}$CO and $^{13}$CO J=3--2 images (Fig.\,\ref{co32waist}), is geometrically thin. 
We measure a width (FWHM) along its minor axis of 0\farcs36 in the less optically-thick $^{13}$CO J=3--2 line. Since the beam has a 
FWHM size $0\farcs37\times0\farcs265$ ($PA=34.9\arcdeg$) and is oriented at about 40\arcdeg~relative to the waist, the latter 
is at best, marginally resolved in a direction perpendicular to its plane, and has a deconvolved minor axis of $\lesssim0\farcs18$ ($\lesssim275$\,AU).

\begin{figure}[!ht]
\resizebox{0.9\textwidth}{!}{\includegraphics{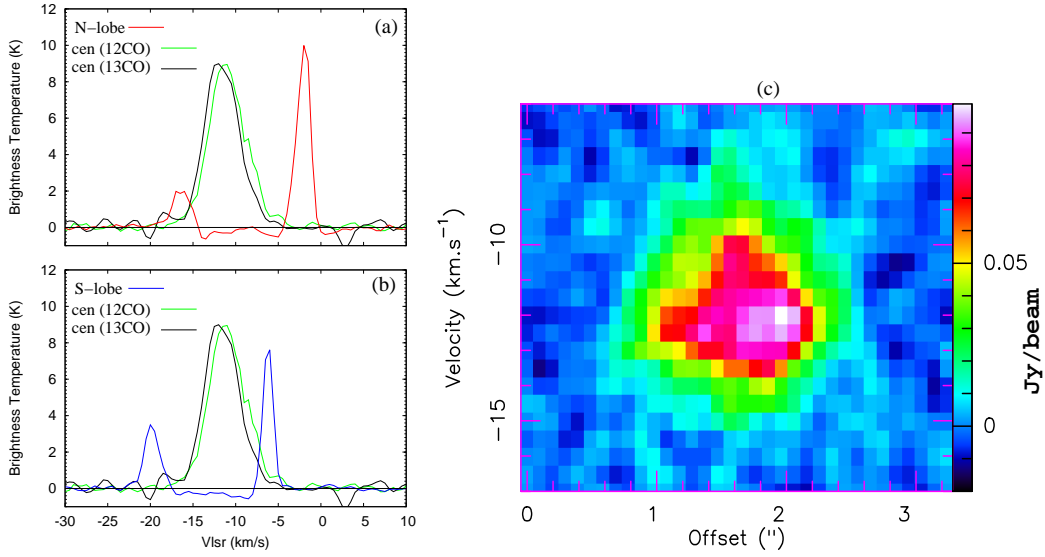}}
\caption{(a,\,b) $^{12}$CO J=3--2 spectra extracted from the northern (red), and southern (blue) lobes of the Boomerang. The $^{12}$CO and 
$^{13}$CO 3--2 spectra from the central waist in green (scaled by 0.5) and black (scaled by 0.95) are shown in each panel for 
comparison. (c) Position-velocity cut of the $^{13}$CO J=3--2 intensity, 
taken along the major axis of the waist of the Boomerang.}
\label{co32-lobe-cen-spec-pv}
\end{figure}

In spectra extracted from a $0.3{''}\times0.2{''}$ region in the waist-center as seen in the J=3--2 $^{12}$CO and $^{13}$CO images, the 
velocity width (FWHM$\sim4.5$\,kms) is significantly smaller than that in the bipolar outflow (Fig.\,\ref{co32-lobe-cen-spec-pv}a,b). A comparison 
of the waist J=3--2 $^{12}$CO line profile with that extracted from our lower angular-resolution J=2--1 $^{12}$CO data (Setal13, Fig.\,3b) spectrum 
shows a very signficant improvement in the isolation of the waist emission from the lobes. 
A position-velocity cut of the $^{13}$CO J=3--2 intensity taken along the major axis of the waist has a morphology that is not a simple ellipse (as 
expected for expansion), and possibly includes a component due to Keplerian rotation (Fig.\,\ref{co32-lobe-cen-spec-pv}c). If true, then the central region of the 
Boomerang resembles the disk in the Red Rectangle and IW\,Car (dubbed disk-prominent post-AGB or dpAGB objects by 
Sahai et al. 2011), that show Keplerian rotation (expansion) in their 
inner 
(outer) regions (Bujarrabal et al. 2016, 2017). Higher angular-mapping with ALMA can easily test this possibility. We note that the velocity-width of the 
Boomerang's waist is the smallest, compared with the velocity-widths of the central torii in a sample of PPNe compiled by Huggins (2012, 2007).

If some (or all) of the material in the Boomerang's waist lies in a rotating disk, we cannot determine its age. If the disk results from a prolonged 
binary interaction, i.e., not a 
common-envelope event (but see \S\,\ref{discus}, where we argue that a common-envelope event is the most plausible scenario), then it is likely 
long-lived, like 
the disks in dpAGB objects, many of which have known central binary stars (de Ruyter et al. 2006, van Winckel et al. 2009, Bujarrabal et 
al. 2013). Assuming that outer regions of the Boomerang's waist are in expansion, we 
roughly estimate an expansion age of 
1925\,yr, by dividing the half-power extent (deconvolved) of the 
$^{13}$CO J=3--2 waist emission ($1\farcs16$ or 1740\,AU) by its half-power velocity-width (FWHM=4.3\,kms). Because the waist emission is not very clearly isolated 
from that arising in and near the base of the lobes, it is likely that the waist size (age) is somewhat smaller than 1740\,AU (1925\,yr). 

The centroid of the $^{13}$CO line is blue-shifted from that of the $^{12}$CO line. This can be explained as the well-known radiative-transfer effect 
in an optically-thick expanding medium where the excitation temperature falls with radius (and may also require a velocity-gradient).
The $^{12}$CO and $^{13}$CO line intensity ratio is quite low, implying that the $^{12}$CO J=3--2 is certainly optically thick. Simple modeling using 
RADEX (Van der Tak et al. 2007) suggest that the kinetic temperature is $\gtrsim$27K, and $N(CO)\sim 10^{17}$\,cm$^{-2}$, resulting in an 
optical depth of the $^{12}$CO J=3--2 line of, $\tau$(3--2)=2.4. 
For $\tau$(3--2) half this value, the model $^{12}$CO J=3--2 brightness temperature is 11.5\,K, significantly lower 
than the observed 16\,K. The number density $n(H2)\gtrsim few\times10^4$ cm$^{-3}$ in order for excitation to be not sub-thermal. If 
the excitation is sub-thermal, then T$_{kin}$ is greater than 
27\,K. For these parameters, the $^{13}$CO/$^{12}$CO ratio is 1/3.3 (the maximum value attained during equilibrium CNO nucleosynthesis), and 
the $^{13}$CO J=3--2 optical depth is 0.5. Of course, if $^{12}$CO J=3--2 is more optically thick, the $^{13}$CO/$^{12}$CO ratio will be lower. Although in 
principle, our Cycle 0 $^{12}$CO J=2--1 data can help constrain the above more stringently, since the beam for the latter is much larger, the waist 
emission cannot be adequately isolated from the bipolar outflow emission.



\subsection{CO\,(J=1--0) Imaging}\label{coj10}

The \codos~J=1--0 mapping (12m mosaic + ACA), together with total-power (TP) data recovers all of the
flux lost in the Cycle 0 data and reveals previously unseen
distant regions. Due to some ISM emission present in the reference OFF position in the TP data, there 
is an absorption spike at -16.7\,\kms.

\begin{figure}[!ht]
\vskip 1.0in
\resizebox{0.8\textwidth}{!}{\includegraphics{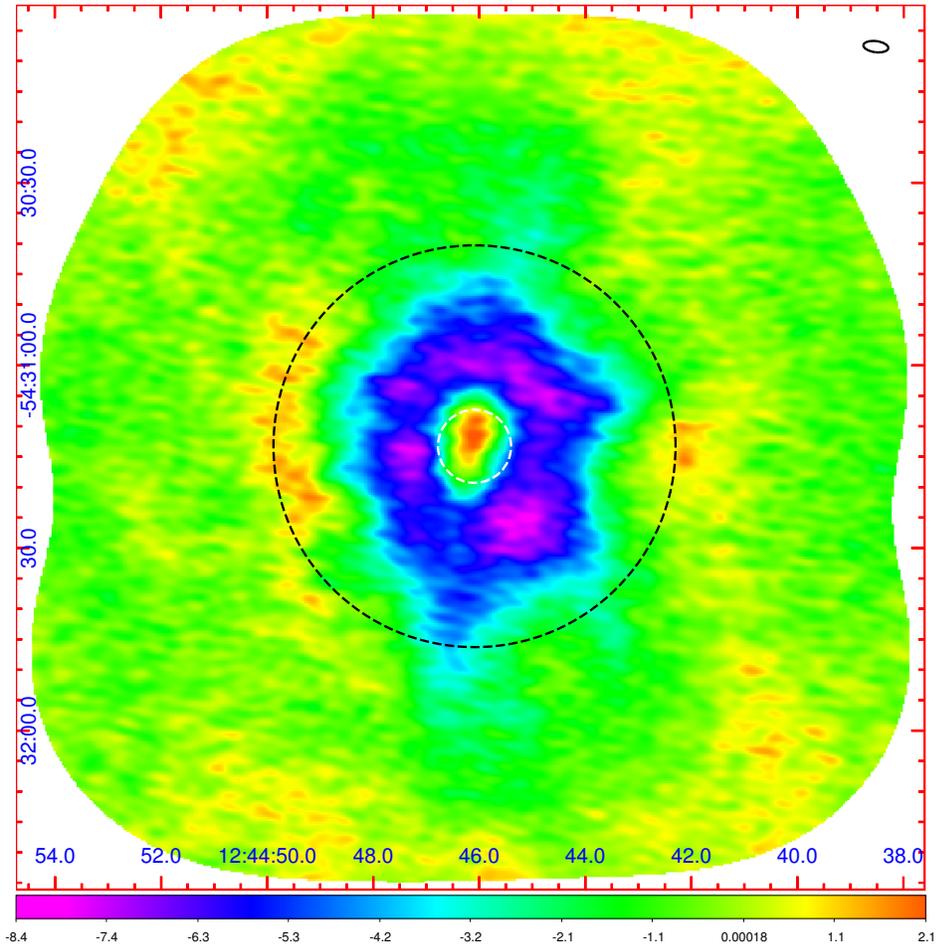}}
\caption{CO J=1--0 moment 0 map of the Boomerang, covering the velocity range, $-171<V_{lsr}\,(\kms)<144$. The black (white) dashed circle denotes 
the size of the SN97 model ultra-cold outflow (inner outflow). The color scale is in units of Jy-km/s beam$^{-1}$; the beam size (FWHM) is shown at the top 
right corner.}
\label{co10mom0}
\end{figure}

\begin{figure}[!ht]
\resizebox{1.0\textwidth}{!}{\includegraphics{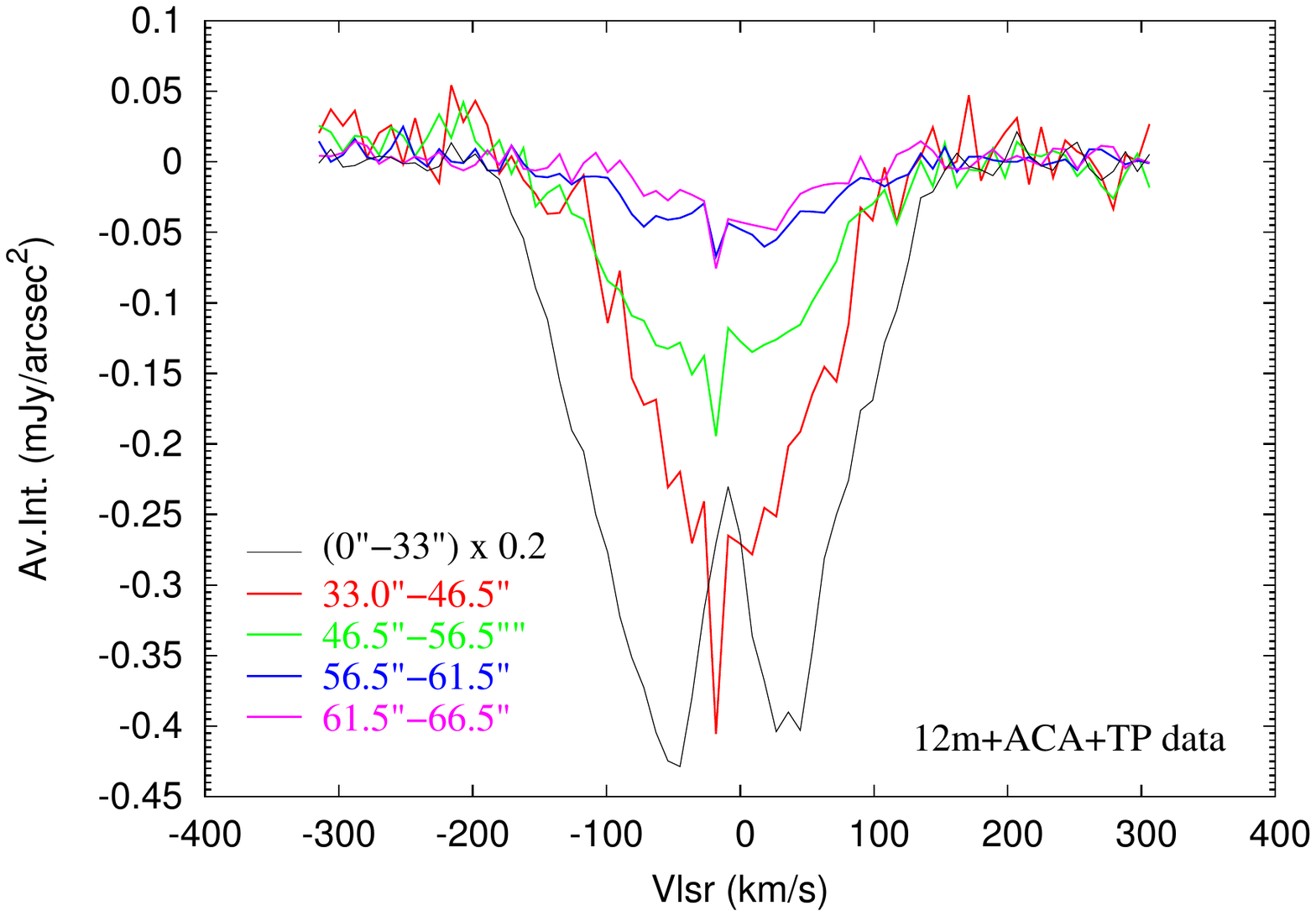}}
\caption{CO J=1--0 spectra extracted from the combined datacube, using annular apertures of increasing average radius.}
\label{co10all_annuli}
\end{figure}

A moment 0 map of the CO J=1--0 absorption+emission, integrated over the velocity range $-171<V_{lsr}\,(\kms)<144$, from 
this dataset (beam $4\farcs1\times1\farcs8$), reveals a mild large-scale asymmetry in the envelope -- the ultra-cold region (seen in magenta/dark 
blue) is slightly more extended N-S (Fig.\,\ref{co10mom0}).  In spectra extracted from annuli with increasing average radii, the 
absorption signal can be detected out to the outermost edge of our map (Fig.\,\ref{co10all_annuli}). The line profile tends to get narrower as the average annulus radius 
increases. 


The true envelope size in CO is difficult to determine. We can only determine a lower-limit from the data because of the rather pathological 
conditions in this object, specifically that the CO from the ultra-cold envelope is in absorption. This signal weakens at larger radii, which can  
result from a warming of the outermost layers and/or a decrease in optical depth such that it becomes optically thin 
in the CO J=1--0 line, causing line excitation temperature to equilibrate with the microwave 
background temperature. Even the decrease in velocity width that we observe at large radii is not necessarily an indicator that we are getting close 
to the edge of the CO envelope, because the above effects will also restrict the total velocity spread of material observed in absorption along 
the line-of-sight.

\begin{figure}[!ht]
\resizebox{1.0\textwidth}{!}{\includegraphics{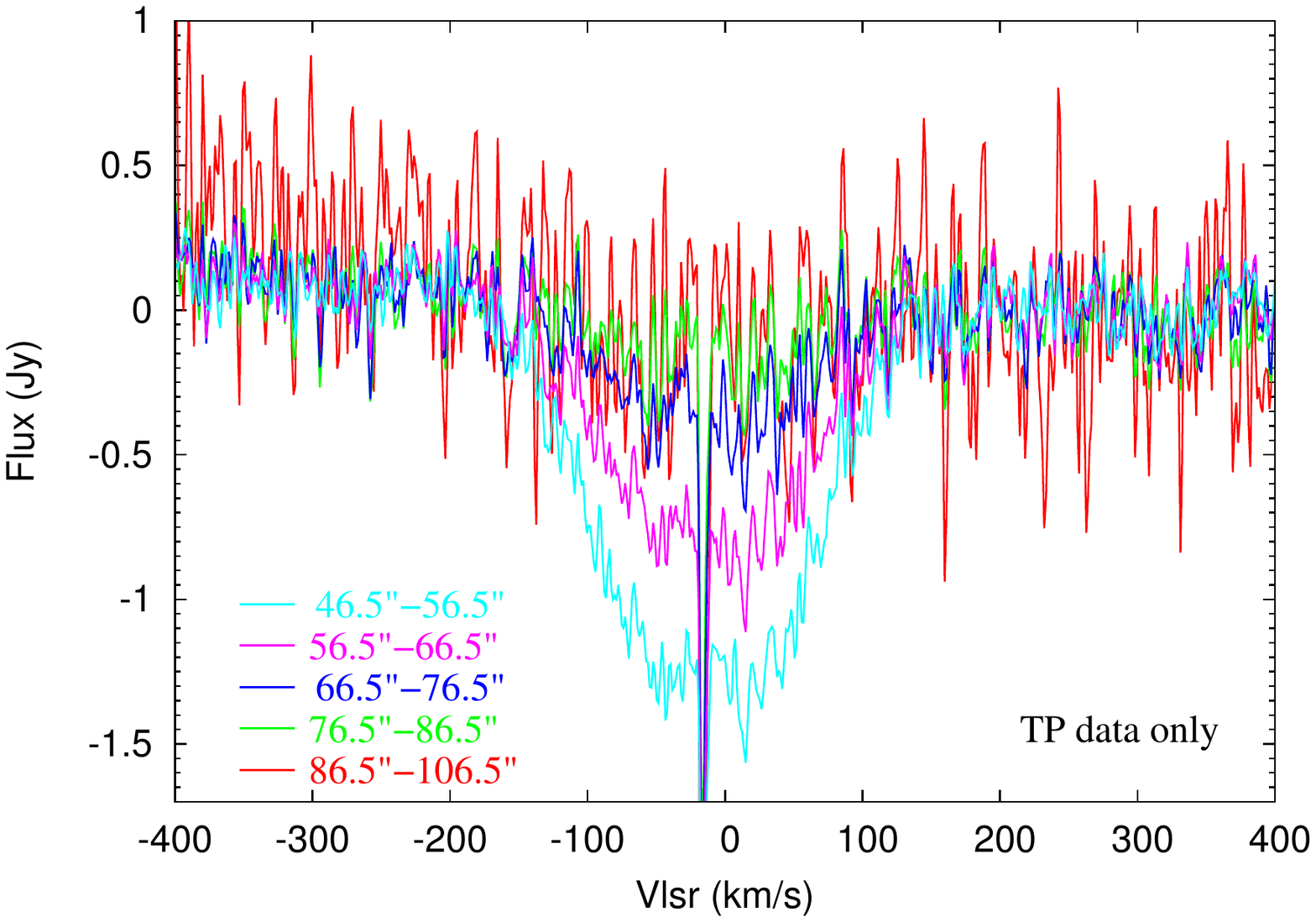}}
\caption{CO J=1--0 spectra extracted from the TP datacube, using annular apertures of increasing average radius.}
\label{co10TP_annuli}
\end{figure}

Using the TP map, that has the largest FOV, we extract spectra from a few annuli, with the outermost one having an outer radius close to the edge of 
the FOV (Fig.\,\ref{co10TP_annuli}). A significant absorption signal can be detected in the $76\farcs5-86\farcs5$ annulus, i.e. at an average 
radius of about $80{''}$. The line profile tends to get narrower as the average annulus radius increases.

\subsubsection{Physical Properties of the Ultra-Cold Outflow}\label{discus}
Our new ALMA CO J=1--0 map of the Boomerang, combining the 12m mosaic with ACA mapping and TP data, shows the presence of the ultra-cold outflow out 
to a radius of at least $80{''}$, significantly larger than that inferred by SN97. 
This finding, together with the discovery of the radially 
increasing expansion velocity that we reported in Setal13, warrants a re-examination of the mass-loss rate and total mass in the ultra-cold outflow. 

In the ultra-cold outflow, since the excitation temperature of the J=1--0 line, $T_{exc}$(1--0) is below the microwave background, 
$T_{bb}=2.725$\,K, virtually all of the CO population is in the J=0 level, and the tangential optical depth in the J=1--0 line, $\tau_{1,0}$, at any 
impact parameter p, is relatively insensitive to the exact value of $T_{exc}$(1--0). We can therefore analytically express $\tau_{1,0}$ as a 
function of a radially-variable mass-loss rate, $\mdot_0 [V(r)/V_0]^a$, and a linearly-increasing outflow velocity,  
$V_{r}=V_0 (r/R_0)$ (see Eqn.\,\ref{tau10eqn}, Appendix). We set $V_0=164$\,\kms~and $R_0=80{''}$.
Since $\tau_{1,0}\ge1$ if $T_{exc}$(1--0)$<T_{bb}$, then setting $\tau_{1,0}\ge1$ at an angular radius, $r=80{''}$, and assuming a constant 
mass-loss rate (i.e., $a=0$), $dM/dt=\mdot _0$, we find that $\mdot_0\ge2.1\times10^{-3}$\,\my. Making conservative assumptions about the 
value of the outflow inner radius, 
$r_{in}=6{''}$, and the fractional CO abundance for a C-rich star (determined by the O abundance, assuming complete association of O into 
CO), $f_{CO}=1.3\times10^{-3}$, the total mass in the ultra-cold outflow is 
$M_{cold}\sim19$\,\ms\,ln$(6{''}/r_{in})$\,$(1.3\times10^{-3}/f_{CO}$)~(Eqn.\,\ref{mcold0eqn}, Appendix). For an O-rich object, $f_{CO}$ is lower 
($0.66\times10^{-4}$: determined by the C abundance, assuming complete association of C into CO), 
requiring an even larger mass (by factor 2) for the slow outflow\footnote{Although it is more likely that the Boomerang is O-rich (see \S\,\ref{findso}), the 
constraints that we derive in this section are intentionally conservative, employing the lower value of $\mdot _0$ based on assuming it to be C-rich.}
Thus, assuming that the Boomerang has evolved from an RGB\footnote{RGB = Red Giant Branch} or AGB star (which implies that its  
main-sequence mass is, $M_{ms}\lesssim8$\,\ms), the inferred values of $M_{cold}$ are unacceptably large.

We therefore conclude that the mass-loss rate in the ultra-cold outflow is not uniform.
The value of $M_{cold}$ can be reduced if we assume that the mass-loss rate in the ultra-cold outflow 
increases with the outflow velocity, 
i.e., $a>0$, provided $\tau_{1,0}>1$ in its inner 
regions, $r\gtrsim\,r_{in}$. 
Using Eqn.\,\ref{mcoldaeqn} (Appendix), we find that $a\gtrsim\,0.9$, since the maximum mass 
that can be ejected in the ultra-fast outflow is $M_{cold}\lesssim7.3$\,\ms, assuming that the 
mass of the Boomerang's central (post-AGB) is comparable to the typical masses for the central stars of 
planetary nebulae (which lie in a relatively
narrow range peaking at $\sim0.61$\,\ms: Gesicki \& Zijlstra 2007), and accounting 
for the (relatively small) mass ejected in the inner outflow ($\sim0.13$\,\ms: SN97).

We can set an 
upper limit on $a$ from the requirement that both the \codos~and \cotres~J=1--0 lines must remain optically thick in the inner regions of the ultra-cold 
outflow in order to keep their excitation temperature there below $T_{bb}$. Specifically, as per the SN97 model, $\tau_{1,0}=5.2$ at 
$r=10{''}$\footnote{note  that $\tau_{1,0}>3.3$, since the $^{13}$CO J=1--0 line also shows 
absorption, and the $^{12}$CO/$^{13}$CO abundance ratio is $\ge3.3$ (SN97)}, 
which implies, using $\tau_{1,0}\propto r^{a-3}$ (Eqn.\,\ref{tau10eqn}, Appendix), that $a\lesssim2.2$, $M_{cold}\gtrsim3.3$\,\ms, 
and $M_{ms}\gtrsim4$\,\ms. In order to maintain $\tau_{1,0}\ge1$ at $r=80{''}$ (see above), Eqn.\,\ref{tau10eqn} (Appendix) implies that 
$\mdot _0$ must increase as $D^{3-a}=D^{(0.8-2.1)}$. Hence, in order to keep $M_{ms}\lesssim8$\,\ms, the distance must be $\lesssim4$\,kpc.

The Boomerang has an anomalously low luminosity, even at its maximum allowed distance of 4\,kpc, of $<2130$\,\ls. For example, 
for a post-AGB object that has 
evolved from a star with $M_{ms}\gtrsim4$\,\ms, post-AGB evolutionary tracks show that the luminosity in the early post-AGB phase is 
$\gtrsim2\times10^4$\,\ls~(e.g., Vassiliadis \& Wood's 1994, Miller Bertolami 2016).

It is possible that the Boomerang is a post-RGB or post-EAGB\footnote{EAGB = Early Asymptotic Giant Branch} 
object, similar to the class of low-luminosity 
($L\sim100-2500$\,\ls), dusty evolved objects 
recently discovered in the Magellanic Clouds (Kamath et al. 2016).  
The Boomerang is not the only evolved star with an anomalously low luminosity; as noted by SN97, the very young PN, M\,1-16, which is similar to the Boomerang 
in being a young bipolar PN and having a very low $^{12}$CO/$^{13}$CO abundance ratio, shares the same low-luminosity issue as the Boomerang 
(Sahai et al. 1994).

\subsection{Discovery of SO Emission}\label{findso}
\begin{figure}[!ht]
\resizebox{0.7\textwidth}{!}{\includegraphics{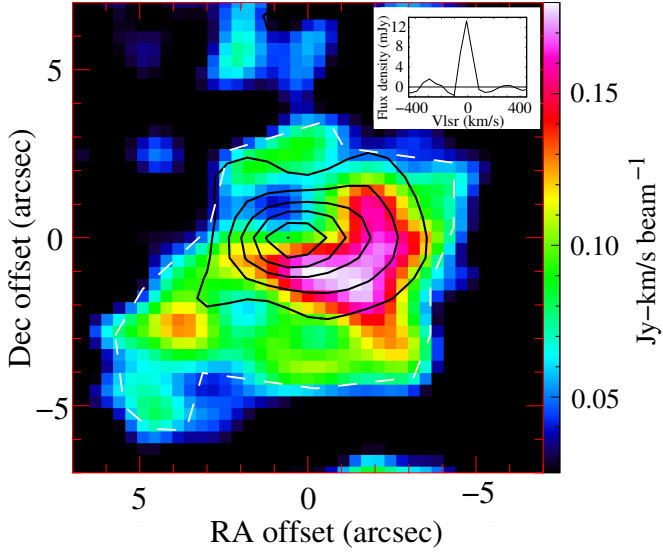}}
\caption{Moment 0 map of the SO N,J=2,3-1,2 line emission (color-scale), overlaid with 
the 92\,GHz continuum emission (contours) towards the Boomerang. 
The minimum contour (step) is 25\% (15\%) of the peak intensity, $3.37\times10^{-4}$\,\jykmsperbeam. The beam for the SO (continuum) data 
is $2\farcs38\times1\farcs81$, $PA$=-87.4\arcdeg ($2\farcs61\times1\farcs97$, $PA$=-89.1\arcdeg). Inset shows the SO line profile, extracted from a polygonal 
aperture (dashed line).}
\label{boom-so}
\end{figure}

We find the unexpected presence of weak SO line (N,J=2,3-1,2 at 99.299\,GHz) emission (Fig.\,\ref{boom-so}) towards 
the central region of the Boomerang. Although the SO map is relatively noisy, the line emission is not centered on the 
3\,mm continuum source but appears to be associated with the 
base of each the N and S lobes. The intrinsic line-width is 79\,\kms~(inset, Fig.\,\ref{boom-so}), 
after deconvolving the limited 
spectral resolution of the continuum spectral window in which the line was found (15.625\,MHz). 

Our detection of SO in the Boomerang suggests that the nebula is O-rich, 
since SO is not expected to be present in the outflows of C-rich AGB stars, and has not been found so far in C-rich objects. Further support for this inference comes 
from the detection of a H$_2$O line in 
the Boomerang (Bujarrabal et al. 2012).  
But the SO and H$_2$O detections are not totally conclusive about the Boomerang's O-rich nature, since H$_2$O has been found in C-rich post-AGB objects (e.g., 
CRL\,618, CRL\,2688, and NGC\,7027: 
Bujarrabal et al. 2012). SO may be produced in the Boomerang as a result of shock-driven chemistry in the 
interaction beteween 
the inner bipolar outflow and the outer ultra-cold outflow, 
possibly involving dust sputtering that releases S locked up as solid MgS in grains (e.g.,  
ISO observations show a strong 30\,\micron~feature in C-rich evolved stars attributed to MgS: 
Hony et al. 2002).

In light of our discovery of SO, we have searched for the presence of additional weak lines. We imaged the entire band 3 SPWs. For the band 3 ``high'' SPW's, 
that cover the  
99.75--103.5\,GHz and 111.75--115.5\,GHz frequency ranges, we made images at 10 channel (4.8826\,MHz) resolution for our 12m-mosaic data (we did not 
include the ACA, as it doesn't add much additional sensitivity). We found no indication of any lines, apart from CO J=1--0.
For the band 3 ``low'' SPW's, that cover the 83.01--88.01\,GHz and 96.01--98.01\,GHz frequency ranges, we made images with single channel spectral 
resolution (15.6\,MHz). We found no indication of any other lines, apart from the SO line mentioned above. The frequency region covered in these 
data, excluding bad channels, does not cover any additional SO line.
 
\subsection{Continuum Imaging and Dust Mass in the Dense Waist}\label{cont_txt}
\begin{figure}[!ht]
\resizebox{0.75\textwidth}{!}{\includegraphics{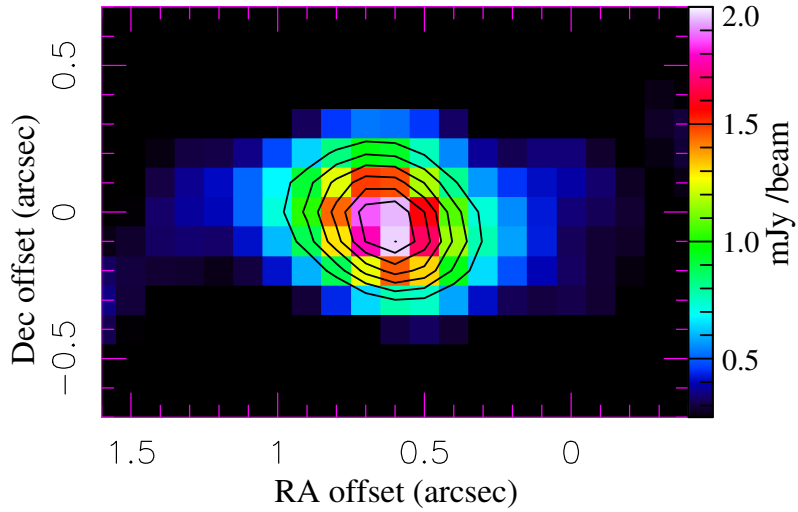}}
\caption{The 0.87\,mm continuum emission from the Boomerang. The continuum peak is located
at (J2000) RA=12:44:46.081, Dec=-54:31:13.38, whereas the phase center (i.e., offset 0,0) is located
at (J2000) RA=12:44:46.01,  Dec=-54:31:13.32. Minimum contour level (step) is 40\% (10\%) of the 
peak, 1.96\,mJy/beam. Beam (FWHM) is $0\farcs43\,\times\,0\farcs35$, at $PA=42.4\arcdeg$.}
\label{cont343}
\end{figure}

We observed the continuum emission at 3.3\,mm and 0.88\,mm towards the Boomerang. At 0.88\,mm, the waist shows an unresolved core with a faint 
extension along the E-W direction; the beam is $0\farcs43\,\times\,0\farcs35$ at $PA=42.4$ (Fig.\,\ref{cont343}). The peak intensity is $2.7\pm0.1$ 
mJy/beam, and the noise in the image is about 0.06 mJy/beam. At 3.3\,mm (beam $2\farcs60\times1\farcs97$ at $PA=-87.4$), there is a bright 
central source that is extended along the E-W direction, with a half-power size that is $3\farcs9\times2\farcs5$ (Fig.\,\ref{boom-so}, contour map). 
In addition, there is 
faint, structured, extended emission N and S of the central source, with peak intensities in the range $0.07-0.09$\,mJy/beam. The noise in 
the 3.3\,mm image is in the range $0.015-0.02$\,mJy/beam.

\begin{figure}[!ht]
\rotatebox{270}{\resizebox{0.6\textwidth}{!}{\includegraphics{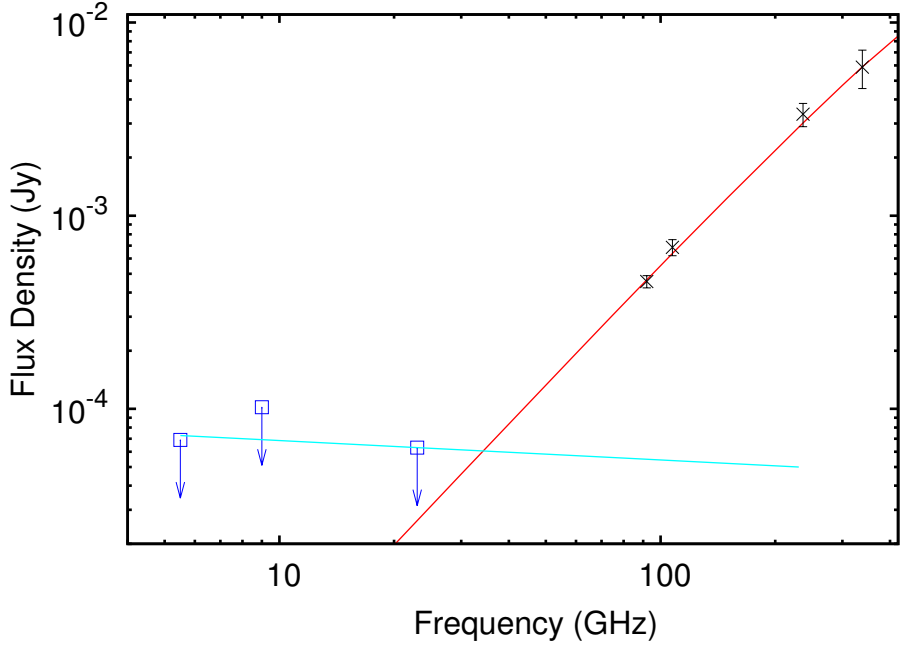}}}
\vskip 0.7in
\caption{The radio to millimeter-wave SED of the Boomerang. Error bars for the ALMA data (black crosses) are 
$\pm1.5\sigma$. The ATCA data show (blue squares) $\pm3\sigma$ upper limits. The ALMA data have been fitted with optically-thin dust emission from 
large grains with temperature 30\,K and a power-law emissivity index, $\beta=0.11$ (red curve). The ATCA upper-limits have been fitted with optically-thin 
free-free emission (cyan curve).}
\label{boom-sed}
\end{figure}

We revise the Setal13 model of the radio-to-(sub)millimeter-wave SED (Table\,\ref{boom-sed-tbl}), using both cycle 0 and cycle 1 
continuum measurements (Fig.\,\ref{boom-sed}), extending the wavelength coverage of the mm-submm SED over a much larger wavelength span, thus 
enabling a more robust estimation of dust properties. The fluxes were extracted 
from images convolved to the same beam as used in the 
lower-resolution Cycle 0 data ($4\farcs1\,\times\,2\farcs9$ at $PA=-33.9$\arcdeg), using an aperture of size $6{''}\times4{''}$ that encompasses  
most of the emission.  The dust 
emission is optically-thin at all wavelengths: e.g., even with the dust temperature, $T_d$, as low as 5--10\,K, the 
peak optical depth at 0.88\,mm, estimated from its peak intensity, is 
about 0.09--0.015. Assuming a power-law 
dust emissivity $\kappa(\nu)\propto\nu^\beta$, and using the fluxes at the maximum and minimum 
frequencies of our observed range (i.e., 338.4 and 92.0\,GHz), we find that $\beta\sim0$ in the Rayleigh-Jeans (R--J) limit. 
However, $\beta$ may be higher if $T_d$, is low enough to make the R--J approximation inaccurate (Setal13). For example, 
if $T_d=15, 20, 25, 30, 35$\,K, then $\beta=0.3, 0.21, 0.16, 0.12, 0.1$.



We follow Setal13's reasoning to constrain $T_d$, using the deconvolved half-power size of the continuum source 
measured at 3.3\,mm, $\sim2\farcs3$ (giving a characteristic radius of $\sim2.6\times10^{16}$\,cm for the dust source). 
The dust radius at a given value $T_d$ is, $r_d=(L_*\,T_{*}^\beta/16\,\pi\,\sigma)^{1/2}\,T_d^{-(4+\beta)/2}$ (e.g., Herman et al. 1986), 
for heating by a central stellar source with luminosity and temperature of $L_*$ and $T_*$ ($\sigma$ is the Stefan-Boltzmann constant). If the dust grains in 
the Boomerang's waist 
are heated directly by 
starlight, then $T_d,\,\beta$=(32\,K,\,0.11). If, as is more likely, there is substantial 
extinction and reddening due to the inner parts of a central disk, and we conservatively assume that only 10\% of the total stellar flux, 
reddened to 900\,K is available for heating, then $T_d,\,\beta$=(19\,K,\,0.22). We conclude that the continuum emission 
source in central region of the Boomerang has $T_d\sim19-32$\,K, $\beta=0.1-0.2$ with a dust 
mass $(4-7)\times10^{-4}$\,\ms. Comparable or larger masses of 
such grains have also been found in the central regions of other PPNe and disk-prominent post-AGB objects (e.g., Sahai et al. 2011). 

The very low value of $\beta$ and 
the lack of observable steepening of the spectral index of the SED toward the largest wavelength observed with ALMA, 3.5\,mm, suggests that the grains  
may have sizes in the cm range. For example, in an extensive laboratory study of grains by Pollack et al. (1994), the 
lowest values of $\beta$ are $0.23-0.87$ for composite\footnote{individual grains contain multiple dust species, with 50\% of void volume} 
grains of radius 3\,cm at temperatures of $700-100$\,K. Compared to composite grains, segregated\footnote{individual grains contain only one dust species} 
grains give lower values of $\beta$ for the same grain-size and temperature, e.g., $\beta=0.31-1.08$ 
for 3\,mm radius grains at temperatures $700-100$\,K, compared to $\beta=0.86-1.34$ for composite grains. Taking into account the 
inverse relationship between $\beta$ and temperature for the laboratory grains, and the relatively low temperature derived for the 
dust grains in the Boomerang, even if the grains in the Boomerang are segregated, it appears likely that the grain sizes are $>$0.3\,cm.

We now discuss and discard two alternative mechanisms for explaining the shallow mm-submm spectral index found above -- spinning dust grains and ferromagnetic 
or ferrimagnetic nanoparticles. 
Spinning dust grains produce cm-wave emission, generally peaking around $\sim40$\,GHz (e.g., Draine \& Lazarian 1998); however under special circumstances the 
emission peak can shift to higher frequencies (Silsbee et al. 2011), and thus could contribute at millimeter wavelengths. Silsbee et al. (2011) 
show plots of the SED due to spinning dust for 6 different astrophysical environments -- amongst these, reflection nebulae (RN) and 
photodissociation regions (PDR) peak at millimeter wavelengths. However, in the RN model the emission peak is at $\sim$85\,GHz and falls relatively 
sharply for higher frequencies (e.g., eqn. 9 of Draine \&  Hensley 2012: DH12), so it does not provide enough flux in the 1.3--0.87\,mm range to explain 
the Boomerang's observed shallow spectral index in this range. The PDR model peaks at a higher frequency ($\sim$170\,GHz) but it is 
characterised by an ionizing radiation field that is a factor 3000 times the intensity of the ambient Galactic starlight, inconsistent with 
the relatively low radiation-intensity environment present in the Boomerang waist region. We conclude that spinning dust grains cannot 
explain the shallow spectral index of the mm-submm SED in the Boomerang.

Emission from ferromagnetic or ferrimagnetic nanoparticles in the cm-submm wavelength range, where these materials have  
enhanced absorptivity, has been 
proposed by Draine \& Hensley (2013) to explain 
the observed strong mm-wave emission from a number of low-metallicity galaxies, including the Small Magellanic Cloud 
(DH12). From Fig. 12 of Draine \& Hensley (2013), 
the spectral index for the mm-submm SED resulting from various types of Fe-bearing nanoplarticles is roughly 
consistent with that of the Boomerang's mm-submm SED. 

We scale the predictions for the emission from different kinds of Fe nanoparticles (radius $a=0.01$\,\micron) 
in models of the SMC SED by DH12, to 
obtain an estimate of the dust mass in Fe-nanoparticles, M$_{d,Fe}$, required to produce the observed (say) 100\,GHz flux in the Boomerang ($\sim0.6$\,mJy).
From Figs.\,3-4 of DH12, we find that F$_{\nu}$(100\,GHz)$\sim0.09$\,mJy (D/62\,kpc)$^2$ (M$_{d,Fe}$/1\,\ms), implying that for the Boomerang, 
M$_{d,Fe}$=$3.9\times10^{-3}$\,\ms. Assuming a cosmic abundance of Fe (Fe/H=$4\times10^{-5}$) and that 100\% of all the available Fe is 
in these small grains, the associated gas mass of 1.8\,\ms~is implausibly large for the waist region of the Boomerang.





\subsection{Optical Polarization}\label{polr}
\begin{figure}[!ht]
\resizebox{1.0\textwidth}{!}{\includegraphics{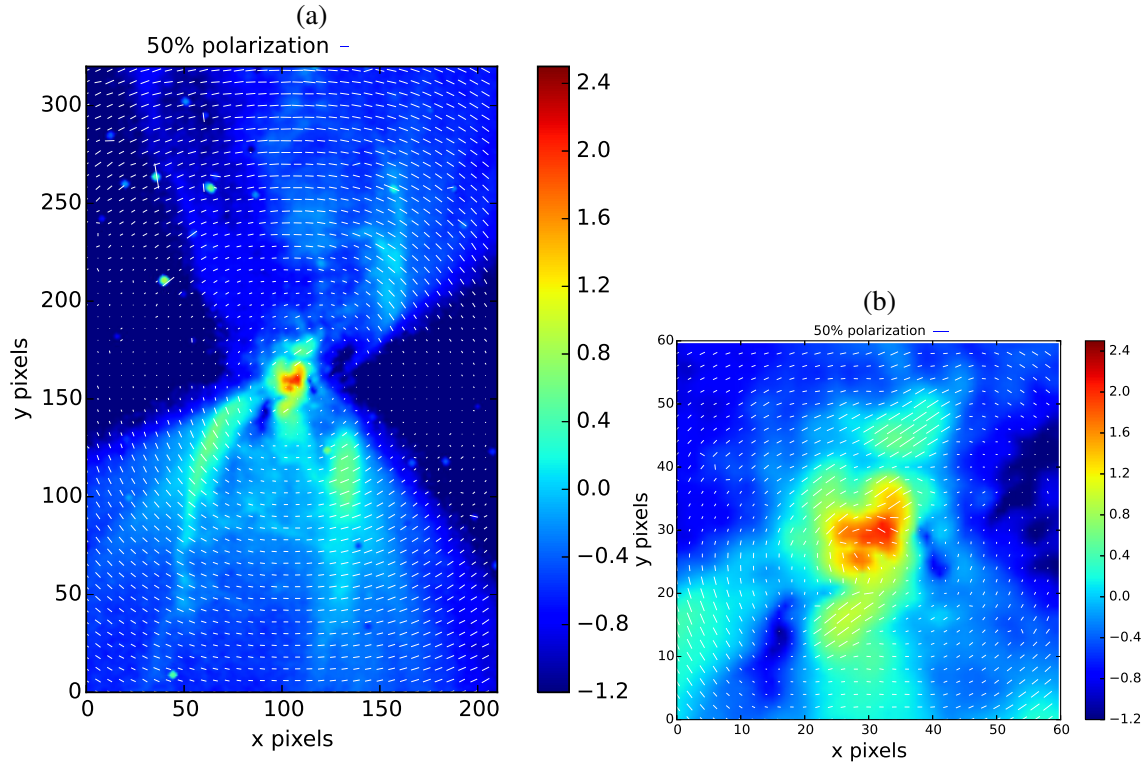}}
\caption{Optical polarization in the Boomerang Nebula as observed with HST/ACS (a) The 0.6\micron~polarized light intensity  
image (log stretch) overlaid with vectors showing the polarization angle and percentage polarization (b) As in $a$, but for the central region.  
Color bars show the log of the intensity, i.e., log10(DN\,pixel$^{-1}$), where 1 DN $=5.016\times10^{-19}$\,\intunit. The angular scale 
is $0\farcs025$\,pixel$^{-1}$.}
\label{polcen}
\end{figure}

The optical polarization pattern (Fig.\,\ref{polcen}a) is generally centro-symmetric around the location of the central star (i.e., polarization 
vectors have an azimuthal orientation) in most 
regions of the nebula except in the waist region, and 
the polarized light fraction is quite high, with values as high as $\sim60$\,\% in the bright walls of the lobes --  
typical of a singly-scattering reflection nebula (e.g., Sahai et al. 1999) with small grains of size $\sim0.1$\micron~(see Table 1 of Jura 1975).
In the dense waist, the percentage polarized is significantly lower most likely due to multiple scattering in an optically-thick medium. 

In the central region, a remarkable S-shaped structure can be seen in 
the polarized intensity image (Fig.\,\ref{polcen}b). The 
polarization vectors in this structure show a striking departure from the centro-symmetric pattern. For example, in the northern and 
southern spurs of this structure, the vectors are oriented along the position angle of the spurs, and the 
polarized fraction is quite high in these regions, about 50\%. Such a high level of polarization can only be explained by single-scattering off dust grains. 
The orientation of the polarization vectors at an intermediate angle between the azimuthal direction and the radial direction suggests the presence of a radial 
polarization component, in addition to the expected azimuthal one. 

Radial polarization has been observed by Woodward et al. (2011) in the optical and near-IR, at scattering angles of $\sim$0\arcdeg-20\arcdeg, from their 
observations of comet C/2007 N3 (Lulin) -- these authors state 
that such ``optical negative branch polarization'' behavior is observed in most comets at small ($\lesssim$25\arcdeg) phase angles. 
Canovas et al. (2015) show, using Mie theory (Mie 1908) that different types of grains 
can rotate the plane of polarization from perpendicular (producing an azimuthal polarization pattern) to parallel (producing a radial polarization pattern). 
They find negative polarizability (leading to radial polarization) for grains 
of different properties when the scattering angle lies in the ranges $\sim$15\arcdeg--60\arcdeg and 140\arcdeg--160\arcdeg (see Fig.\,1b of Canovas et al. 
2015).

Based on the results of these studies, we hypothesize that the polarization pattern observed in the spur structures in the Boomerang 
is a result of optical negative branch polarization due to a fortuitous location of the spurs within the lobes, such that some fraction of the 
material within them provides scattering angles for which the polarizability is negative. For example, if some sizable fraction of the 
southern and northern spurs lie along a common axis with inclination, $\theta_{spur}$ relative to the line-of-sight ($los$), then for (say)  
$\theta_{spur}=20\arcdeg-40\arcdeg$, the scattering angles of these regions in the southern (northern) spur are $20\arcdeg-40\arcdeg$ 
($160\arcdeg-140\arcdeg$), i.e., within the negative polarizability angle ranges in the Canovas et al. (2015) models. 
Material in the spurs that subtend scattering angles outside these ranges would have positive polarizability, and the 
combination of radial and azimuthal polarization vectors could then produce the observed intermediate orientation. Note that for our 
hypothesis to work, it is essential that the spur material lie within a compact region along the $los$, implying that the S-shaped 
structure is filamentary. We speculate that this structure may be due to a precessing jet.

\section{Ejection Time-Scales and Binary Interaction Models}\label{discus}
We estimate a lower limit on the expansion age of the ultra-cold outflow of 3480\,yr (from $R_0/V_0$). Comparing this with the ages of the bipolar lobes, 
and the central waist, we find that the ultra-cold outflow is the oldest, followed by the waist\footnote{assuming its outer region is expanding}  
($\sim1925$\,yr), and the youngest is the fast bipolar outflow(s) ($\le1050$\,yr) 
that interacts with the ultra-cold outflow to produce the N- and S-lobes. Thus the 
temporal sequence of waist-formation followed by the ejection of a collimated fast outflow, in the Boomerang,  
is similar to that derived by Huggins (2007) for a small sample of 
similar objects (late AGB stars, PPNe and young PNe). Huggins shows that this sequence naturally favors the class of models in which a companion interacts 
directly with the central star. These include two subclasses of models: (A) the build-up of a torus 
enhances the accretion rate in a disk around a companion that then drives jet-like outflows, or 
(B) both spin-up and ejection of the stellar envelope of the primary occur during a common-envelope (CE) phase as the companion spirals in to the 
center of the AGB star. 

We note that the time periods between the CSE ejection, torus ejection and high-velocity outflow formation 
in the Boomerang (1555\,yr and $\ge875$\,yr) are much larger than the corresponding durations for the ``water-fountain'' PPN, IRAS\,16342-3814, in which these are  
about $300$\,yr and $50$\,yr (Sahai et al. 2017). 
An inspection of the 9 objects in Huggin's study (see his Table 1) reveal that only one has a similarly long ``jet-lag'' (i.e., torus age minus the fast outflow age) 
(KjPn\,8: 1660\,yr), as the Boomerang. Six of the remaining eight have significantly shorter jet-lags ($\sim270-390$\,yr), and 
two have intermediate jet-lags ($\sim510-690$\,yr). Quantitative models that can provide estimates of the ``jet-lag'' have the potential 
of discriminating between subclass A and B models, and constraining the relevant physics.




In the case of the Boomerang, we show that model B is needed, in order to explain the extreme properties of the ultra-cold outflow. The 
very high mass-loss rate, radially-increasing and high expansion velocity of the ultra-cold outflow, 
coupled with the low luminosity of the Boomerang, imply that the standard model for a dusty, molecular outflow from an evolved star, i.e., one driven 
by radiation pressure on dust, is not applicable. The ultra-cold outflow's mass-loss rate of $>10^{-3}$\,\my~is 
orders of magnitude above typical values for the RGB or EAGB ($\lesssim10^{-6}$\,\my~for $L\lesssim2500$\,\ls, Groenewegen 
2012). 
The kinetic energy of 
the ultra-cold outflow is very high, $KE_{cold}>4.8\times10^{47}$\,\engyunit~(using Eqn.\,\ref{keeqn}, Appendix). 

We propose that the most likely 
source of this energy is the gravitational energy released via binary interaction in a common-envelope event (CEE). The latter is given by the 
difference in orbital energies of the binary before and after the interaction, i.e.,  
\begin{equation}
E_{CEE}=\frac{G\,M_2\,M_{1,c}}{2\,a_f}-\frac{G\,M_2\,M_1}{2\,a_i}
\end{equation}
which can be rewritten as, 
\begin{equation}\label{eceeeqn}
E_{CEE}=\frac{G\,M_2\,M_{1,c}}{2\,a_f}\,\eta(M_1,M_{1,c},a_f,a_i)
\end{equation}
where $\eta(M_1,M_{1,c},a_f,a_i)=1-(M_1/M_{1,c})(a_f/a_i)$, $M_1$ ($M_{1,c}$) is the initial (final) mass of the 
primary, $M_2$ is the companion mass, and 
$a_i$ ($a_f$) is the initial (final) semi-major axis of the binary. 

The released energy must be equal to or exceed the kinetic energy in the ultra-cold outflow plus the 
energy required to unbind the envelope, $\left|E_{bind}\right|-\Delta E_{therm}-\Delta E_{other}$, where 
$E_{bind}$ is the binding energy of the envelope, $\Delta E_{therm}$ is the 
thermal energy of the envelope, and $\Delta E_{other}$ can include contributions from recombination, nuclear fusion\footnote{due to a 
non-compact companion, e.g., a main-sequence star, filling its Roche lobe and causing H-rich materal to fall onto primary's core}, and 
accretion\footnote{onto the companion during its in-spiral} (Ivanova et al. 2013). 
Since estimating the value of $\left|E_{bind}\right|-\Delta E_{therm}-\Delta E_{other}$ depends on several uncertain parameters (Ivanova et al. 
2013), we first set it to equal to zero. Hence,    
\begin{equation}
E_{CEE}>KE_{cold},
\end{equation}
from which we can derive an upper limit on the separation of the final binary.

Taking $M_1=4$\,\ms, and typical values for $M_{1,c}=0.6$\,\ms, $M_2=1$\,\ms, and setting $\eta\sim1$ (since $a_i>>a_f$), we 
find that $2\,a_f<4.7$\,\rsun. However, because 
the radius of the Boomerang's central star, estimated to be $\sim16$\,\rsun, from its effective temperature, 
$T_{eff}=6000$\,K and $L\sim300$\,\ls, is significantly larger than 
$2\,a_f$, it is likely that the companion merged with the primary after the in-spiral.

The Boomerang Nebula thus appears to have resulted from a strongly interacting binary system, in which a very significant fraction of the 
primary star's envelope was ejected while it was on the RGB or EAGB, as the companion spiralled in towards the latter's core. This 
ejection contains material covering a large range of velocities (upto a maximum of about 164\,\kms), and produces the ultra-cold outflow.  
At the end of the in-spiralling event, the companion merges with the primary. During the in-spiral, a 
large disk is created around the central star. The disk powers a fast collimated outflow that 
interacts with the ultra-cold outflow to produce the central bipolar nebula. In this scenario, the age of the waist must be less than that of the 
ultra-cold outflow, i.e., $\lesssim 3480$\,yr. 

A potential difficulty with the ultra-cold outflow resulting from CEE is that the former's morphology does not appear to be concentrated 
towards the equatorial plane, as has been found in numerical simulations of CEE (e.g., Iaconi et al. 2017, and references therein). However, a 
detailed 
inspection of the simulation in Iaconi et al. (2017), shows that the mass distribution of the ejecta appears to become more isotropic with time (see 
their Fig. 5, right-hand panel). In Fig. 8 of Iaconi et al. (2017), which shows the masses ejected in 3 pairs of pyramids centered at the centre
of the computational cube and whose bases are the six faces of this cube, the masses in the x- and y-pyramids (these encompass the equatorial plane)  
appear to approach that in the z-pyramid (which encompasses the polar direction) at 2000 days after the onset of the interaction, with only a factor 
3 difference between the mean mass in the x- and y- pyramids, compared to the z-pyramid. 

Given that we are observing the Boomerang ultra-cold outflow at an age that is a factor 
$\sim650$ larger than the 2000 day timespan of Iaconi et al.'s simulation, it is not implausible that the lower-density polar regions seen on small 
scales in the simulation get filled in with time (e.g., small perturbations in the velocity vectors away from radial would allow material to move 
towards the axis). Furthermore, the Iaconi et al. (2017) simulation assumes a much less massive primary (0.88\,\ms~RGB primary) than the Boomerang's; 
its possible that for a more massive RGB star, the CEE ejection is more isotropic. 

New CEE simulations, with more massive primary stars, that can reproduce the relatively well-defined properties of the Boomerang Nebula, will 
be very useful in improving our understanding of an important channel for binary star evolution.




\section{Conclusions}\label{conclude}
We have obtained ALMA maps of millmeter-wave line and continuum emission at the highest angular resolution to-date of the 
Boomerang Nebula, the coldest known object in the Universe.\\
\begin{enumerate}
\item The high-resolution \codos~J=3--2 imaging of the inner outflow of the Boomerang reveals a detailed correspondence between the 
molecular gas and the dust as seen in scattered light via optical HST imaging. Both the molecular-line and the optical images shows that the N-lobe has a  
complex multipolar structure, whereas the S-lobe has a relatively simple (largely cylindrical) geometry. 
\item A dense central waist of size (FWHM) $\sim$1740\,AU$\,\times275$\,AU, separates the N- and S-lobes, and is expanding much 
more slowly than the lobes. The velocity-width of the \codos~and \cotres~J=3--2 line profiles is the smallest seen towards 
a sample of PPNe, and similar to that seen in the rotating disks of post-AGB objects like the Red Rectangle. A position-velocity cut of 
the \cotres~J=3--2 taken along the major axis of the waist shows more complex kinematics than simple expansion, possibly 
due to the presence of a Keplerian rotational component.
\item The ultra-cold outflow in the Boomerang Nebula extends to a radius of at least $80{''}$ (120,000\,AU). 
The mass-loss rate in the ultra-cold outflow ($\ge2.1\times10^{-3}$\,\my~at $r=80{''}$) is 
not uniform, but increases with radius, similar to its expansion velocity, which also increases with radius (Setal13).
The mass in the ultra-cold outflow is $\gtrsim3.3$\,\ms, and the mass of the Boomerang's main-sequence progenitor is 
$\gtrsim4$\,\ms.
\item The expansion age of the ultra-cold outflow is $>3480$\,yr. The central waist (assuming its outer region is expanding) and fast bipolar outflow are 
younger, with expansion ages of $\sim1925$\,yr and $\le1050$\,yr. The ``jet-lag'' (i.e., torus age minus the fast outflow age) 
in the Boomerang lies within (and near the upper end of) the range of values found for other AGB or post-AGB objects with jets and 
torii, and suppports models in which the primary star interacts directly with a binary companion.
\item The relatively shallow slope of the submm-mm SED for $\lambda=1.3-2.6$\,mm found in our 
previous study, applies  
over a much broader wavelength window, $\lambda=0.87-3.5$\,mm. We infer that the grains are cold ($19-32$\,K), and 
that the dust absorption absorption power-law exponent ($\kappa\propto\nu^\beta$) is, $\beta=0.1-0.2$, implying the 
presence of very large (few mm to cm-sized) grains. Alternative models for producing the low value of $\beta$ such 
as spinning grains or Fe-containing nanoparticles are very unlikely.
\item The nebula is highly polarized at optical wavelengths, and the polarization pattern is generally centro-symmetric around the location of the central star 
in most regions of the nebula, typical of a singly-scattering reflection nebula with small grains ($\sim0.1$\micron).
In the dense waist, the percentage polarized is significantly lower, most likely due to multiple scattering in an optically-thick medium.
\item In the central region, an S-shaped structure is seen in 
the polarized intensity image, and the polarization vectors in this structure show a striking departure from the centro-symmetric pattern. 
We hypothesize that this pattern results from the contribution of a radial 
polarization component due to scattering by dust grains at angles for which the polarizability is negative.
\item We find the presence of SO in the central region of the Boomerang. This result, taken together with a previous detection of H$_2$O,
suggests that the Boomerang Nebula is O-rich.
\item The Boomerang Nebula was most likely produced by a common-envelope binary interaction, while the primary was an RGB or 
early-AGB star, with the companion spiralling into and finally merging with the primary's core, and ejecting 
the primary's envelope that now forms the ultra-cold outflow.
\end{enumerate}

\section{Acknowledgments}  We thank an anonymous referee for his/her remarks that have helped improve our paper. We thank 
Orsola De Marco for reviewing and providing comments on our discussion of the in-spiral energetics, and Eric Blackman for his 
thoughts on the possibility of isotropic ejection during CEE. 
We thank Misty Cracraft (STScI) for providing us the calibrated 
HST polarization data on the Boomerang in digital form. This paper makes use of the following ALMA data: 
ADS/JAO.ALMA\#2012.0.00510.S. ALMA is a partnership of ESO (representing its 
member states), NSF (USA) and NINS (Japan), together with NRC (Canada) and 
NSC and ASIAA (Taiwan), in cooperation with the Republic of Chile. The Joint 
ALMA Observatory is operated by ESO, AUI/NRAO and NAOJ. The National Radio Astronomy Observatory is a facility of the National Science Foundation
operated under cooperative agreement by Associated Universities, Inc. RS's contribution to the research described here
was carried out at JPL, California Institute of Technology, under a contract with NASA. WV acknowledge supports from ERC
consolidator grant 614264.

\section{appendix}
\subsection{Analytic Formulation for CO J=1-0 Optical Depth in an Ultra-Cold Outflow}
Using the formulation of Morris (1975), we derive the radiative-contact length, $\delta z(p,v_p)$, which is the distance along any line of sight  
(at a given impact parameter, $p$, relative to the center of the ultra-cold outflow), over which molecules are in radiative contact for a given 
line-width (due to microtubulence and thermal-broadening), $\Delta V$. Thus, 
\begin{equation}
\delta z=\Delta V/(dv_p/dz).
\end{equation}\label{vexpeqn}
Assuming a linearly increasing expansion velocity, 
\begin{equation}
V_{r}=V_0 (r/R_0),
\end{equation}
we find 
\begin{equation}
\delta z=R_0 (\Delta V/V_0).
\end{equation}
The opacity per unit length, for a 
line between rotational energy levels J and J=1, is a product of an absorption coefficient, $\alpha_{J+1,J}(r)$, the fractional abundance of the 
molecule, $f_{mol}$, and the total particle density, $N(r)$. Hence the optical depth is
\begin{equation}
\tau_{J+1,J}=\alpha_{J+1,J}(r) f_{mol} N(r) \delta z.
\end{equation}
Since 
\begin{equation}\label{taueqn}
\alpha_{J+1,J}(r)=\frac{8\,\pi^3\,\mu{_0}^2}{3\,h\,\Delta V} (J+1) (n_J-n_{J+1}),
\end{equation}
then using 
\begin{equation}\label{denseqn}
N(r)=\frac{\mdot(r)}{4\,\pi\,m_{H_2}\,r^2\,V(r)},
\end{equation}
and assuming a mass-loss rate that varies with the outflow velocity as
\begin{equation}\label{mdoteqn}
\mdot(r)=\mdot_0 [V(r)/V_0]^a,
\end{equation}
we get
\begin{equation}\label{taueqn2}
\tau_{J+1,J}=(\frac{8\,\pi^3\,\mu{_0}^2\,f_{mol}}{3\,h\,V_0}) (\frac{\mdot_0}{4\,\pi\,m_{H_2}\,V_0})  (J+1) (n_J-n_{J+1}) (R_0/r)^{2-a}/r.
\end{equation}
Setting  $T_{exc}$(1--0)$<T_{bb}$ ($T_{bb}=2.725$\,K, the microwave background temperature), 
$n_1=n_0\,{\rm exp} \frac{-5.53\,K}{T_{exc}(1-0)}$, hence $n_1<<n_0$, and 
$n_0=1$, (i.e., virtually all of the CO population is in the J=0 level), and $\tau_{1,0}$ is independent of $T_{exc}$, i.e., 
\begin{equation}\label{tau10eqn}
\tau_{1,0}=(\frac{8\,\pi^3\,\mu{_0}^2\,f_{mol}}{3\,h\,V_0}) (\frac{\mdot_0}{4\,\pi\,m_{H_2}\,V_0}) (R_0/r)^{2-a} n_0/r.
\end{equation}
The total mass of the ultra-cold outflow is given by, 
\begin{equation}\label{mcold0eqn}
M_{cold}=\frac{\mdot_0\,R_0}{V_0}\,{\rm ln}(r_{ou}/r_{in}),\,\,\,\,a=0, 
\end{equation}
and
\begin{equation}\label{mcoldaeqn}
M_{cold}=\frac{\mdot_0\,R_0}{a\,V_0}\,[(r_{ou}/R_0)^a-(r_{in}/R_0)^a],\,\,\,\,\,a\ne0,
\end{equation}
where $r_{ou}$ ($r_{in}$) is the outer (inner) radius of the ultra-cold outflow.
The kinetic energy in the ultra-cold outflow is, 
\begin{equation}\label{keeqn}
KE_{cold}=\frac{\mdot_0\,R_0\,V_0}{2a+4}\,[1-(r_{in}/r_{ou})^{a+2}], 
\end{equation}

\newpage
\subsection{Common-Envelope Ejection and Binding Energy}
We re-estimate the final binary separation, $2\,a_f$, accounting for the binding and thermal energy of the envelope, i.e., 
\begin{equation}\label{ebindeqn}
E_{CEE} = \left|E_{bind}\right|-\Delta E_{therm}-\Delta E_{other}+KE_{cold}.
\end{equation}
We estimate $E_{bind}$ using  Eqn. 13 of De Marco et al. 2011 (hereafter DeMetal11), 
\begin{equation}
E_{bind}=-G\,M_e\,(M_e/2+M_{1,c}) / (\lambda\,R),
\end{equation}
where $\lambda$~is a parameter with a value of order unity\footnote{see Ivanova et al. (2013) for a detailed discussion of this parameter}, R is 
the radius of the primary's Roche lobe at the start of the interaction, and $M_e$ is the mass of ejected envelope (which we assume ro be equal to the 
mass of the ultra-cold outflow). The thermal  
energy term, $\Delta E_{therm}$ is half the value of $E_{bind}$, but with the opposite sign (DeMetal11). We ignore the very 
uncertain term, $\Delta E_{other}$, which includes contributions from recombination, nuclear fusion, and 
accretion. The recombination contribution is about 
$E_{rec}\sim(1.5-3)\times10^{46}$\,erg for a RGB star of mass 1.6\,\ms~(see tabulated values of $E_{rec}$ in Tables 1 \& 2 in Ivanova et al. 
2015), and would be larger for the more massive progenitor of the Boomerang.

As before, we take $M_{1,c}=0.6$\,\ms, and $M_e=M_{cold}=3.3$\,\ms. 
From Table 2 of DeMetal11, we find $\lambda\sim0.22-0.12$ if the primary is on the RGB, and 
$\lambda\sim0.33-0.39$ if it is on the AGB, for $M_1=3-5$\,\ms. We take $R\sim100$\,\rsun~for a post-RGB object (Table 3 of DeMetal11) and 
$R\sim200$\,\rsun~for a post-AGB object (Eqn. 20 of DeMetal11, with $M_1=4$\,\ms). 

Hence, for a post-RGB (post-AGB) object, $\left|E_{bind}\right|-\Delta E_{therm}\sim(0.64-1.2)\times10^{48}$ ($\sim(0.18-0.21)\times10^{48}$)\,\engyunit, 
and $E_{CEE}\sim (1.1-1.6)\times10^{48}$ ($\sim (0.66-0.69)\times10^{48}$)\,\engyunit. Using Eqn\,\ref{eceeeqn} for $E_{CEE}$, we find that since 
$a_f<<a_i$, $\eta\sim1$, and $2\,a_f\sim1.4-2.0\,(3.3-3.4)$\,\rsun~for a post-RGB (post-AGB) object. We note that (i) $M_{cold}=3.3$\,\ms~is the minimum value 
of the ejected mass; for larger values, $a_f$ will be smaller, and (ii)  depending on the relative contribution of $\Delta E_{other}$ to the RHS in 
Eqn\,\ref{ebindeqn} above, the value of $a_f$ will be larger.

\clearpage
\scriptsize
\begin{longtable}{p{0.4in}p{0.8in}p{0.4in}p{0.4in}p{0.4in}p{0.5in}p{1.0in}p{0.9in}p{0.4in}}
\caption[]{Log of Observations}\\
\hline \\[-2ex]
   \multicolumn{1}{l}{\textbf{Band}} &
   \multicolumn{1}{l}{\textbf{Lines}} &
   \multicolumn{1}{l}{\textbf{Freq}\footnotemark[1]} &
   \multicolumn{1}{l}{\textbf{Width}} &
   \multicolumn{1}{l}{\textbf{$\delta\,V$/$\delta\,\nu$\footnotemark[2]}} &
   \multicolumn{1}{l}{\textbf{Array}} &
   \multicolumn{1}{l}{\textbf{Beam}}  &
   \multicolumn{1}{l}{\textbf{Date}} &
   \multicolumn{1}{l}{\textbf{Time\footnotemark[3]}} \\[0.5ex]
   \multicolumn{1}{l}{\textbf{}} &
   \multicolumn{1}{l}{\textbf{}} &
   \multicolumn{1}{l}{\textbf{GHz}} &
   \multicolumn{1}{l}{\textbf{GHz}} &
   \multicolumn{1}{l}{\textbf{}} & 
   \multicolumn{1}{l}{\textbf{}} &
   \multicolumn{1}{l}{\textbf{${''}\times{''}$,\,(PA)$\arcdeg$}} &
   \multicolumn{1}{l}{\textbf{dd-mm-yyyy}} &
   \multicolumn{1}{l}{\textbf{min}} \\[0.5ex] \hline
   \\[-1.8ex]
\endfirsthead

\multicolumn{3}{c}{{\tablename} \thetable{} -- Continued} \\[0.5ex]
\caption[]{Log of Observations}\\
\hline \\[-2ex]
   \multicolumn{1}{l}{\textbf{Band}} &
   \multicolumn{1}{l}{\textbf{Lines}} &
   \multicolumn{1}{l}{\textbf{Freq}\footnotemark[1]} &
   \multicolumn{1}{l}{\textbf{Width}} &
   \multicolumn{1}{l}{\textbf{$\delta\,V$/$\delta\,\nu$\footnotemark[2]}} &
   \multicolumn{1}{l}{\textbf{Array}} &
   \multicolumn{1}{l}{\textbf{Beam}}  &
   \multicolumn{1}{l}{\textbf{Date}} &
   \multicolumn{1}{l}{\textbf{Time\footnotemark[3]}} \\[0.5ex]
   \multicolumn{1}{l}{\textbf{}} &
   \multicolumn{1}{l}{\textbf{}} &
   \multicolumn{1}{l}{\textbf{GHz}} &
   \multicolumn{1}{l}{\textbf{GHz}} &
   \multicolumn{1}{l}{\textbf{}} & 
   \multicolumn{1}{l}{\textbf{}} &
   \multicolumn{1}{l}{\textbf{${''}\times{''}$,\,(PA)$\arcdeg$}} &
   \multicolumn{1}{l}{\textbf{dd-mm-yyyy}} &
   \multicolumn{1}{l}{\textbf{min}} \\[0.5ex] \hline
   \\[-1.8ex]
\endhead
3  & $^{12}$CO\,(1--0) & 115.271 & 1.875 & 1.3  & 12m     & $2.26\times1.29$,\,70.7 &  17-12-2013    &  52  \\
   & ...\footnotemark[4]& ... & ... & ...   & ACA       & $12.7\times8.4$,\,84.6  &  20/21-03-2014 &  192  \\
   & ...           & ... & ... & ...   & TP             & $57.4\times57.4$        &  25/26-04-2015 &  160  \\
   & SO\,(2,3-1,2) & 99.2999 & 2.0    & 47    & 12m     & $2.38\times1.81$,\,-87.4&  01-12-2013    & 72  \\
   & cont          & 85.01   & ...    & 31.25 & ...     & $2.60\times1.97$,\,-89.1\footnotemark[5] &  ...   & ...  \\
   & ...           & 87.01   & ...    & ... &  ...      & ... &  ...  &  ...  \\
   & ...           & 97.01   & ...    & ... & ...       & ... &  ...  &  ...  \\
   & ....          & 99.01   & ...    & ... & ...       & ... &  ...  &  ...  \\
7  & $^{12}$CO\,(3--2) & 345.796  & 1.875 & 0.5   & ...   & $0.37\times0.25$,\,34.4 & 14/15-06-2014  &  96  \\
   & $^{13}$CO\,(3--2) & 330.588  & ...   & ...   & ...   & $0.37\times0.27$,\,34.9 & ...  &  ...  \\
   & cont              & 331.26   & ...   & 0.49  & ...   & $0.43\times0.35$,\,42.4 &  ...  &  ...  \\
   & ...               & 333.15   & ...   & ...   & ...   & ... &  ...  &  ...  \\
   & ...               & 343.35   & ...   & ...   & ...   & ... &  ...  &  ...  \\
   & ...               & 345.25   & ...   & ...   & ...   & ... &  ...  &  ...  \\  
\hline
\label{obslog}
\end{longtable}
\footnotetext[1]{Line center frequency, or center-frequency for continuum band}
\footnotetext[2]{Velocity width ($\kms$) per channel for line, frequency width (MHz) per channel for continuum}
\footnotetext[3]{Total observations time (includes on-source integration time and other overheads)}
\footnotetext[4]{``...'' indicates ``same as'' above everywhere in the table}
\footnotetext[5]{Beam parameters for continuum datacube combining 4 bands at 85.01, 87.01, 97.01, and 99.01\,GHz, with center frequency $\nu=92.02$\,GHz}


\clearpage

\begin{table}[!t]
\caption{Boomerang Nebula Continuum Emission}
\label{boom-sed-tbl}
\begin{tabular}{lllll}
\hline
$\nu$  & Peak\,Intensity\footnotemark[1]&    rms & Flux\,Density\footnotemark[2] & Comment\\
GHz    & mJy\,beam$^{-1}$ &  mJy\,beam$^{-1}$  & mJy        &        \\
\hline
85.01   & 0.409   & 0.039   & 0.429   & Individual\,SPW    \\
87.01   & 0.425   & 0.036   & 0.417   & Individual\,SPW    \\
97.08   & 0.521   & 0.047   & 0.536   & Individual\,SPW    \\
99.01   & 0.615   & 0.050   & 0.599   & Individual\,SPW    \\
92.02   & 0.452   & 0.022   & 0.456   & All\,SPW    \\
107.47  & 0.640   & 0.041   & 0.686   & Cycle 0    \\
107.52  & 0.808   & 0.077   & 0.821   & All\,SPW   \\
236.40  & 3.650   & 0.334   & 3.351   & Cycle 0    \\
331.39  & 6.857   & 0.946   & 5.331   & Individual\,SPW    \\
333.13  & 5.911   & 0.732   & 5.232   & Individual\,SPW    \\
343.32  & 6.113   & 0.798   & 5.313   & Individual\,SPW    \\
345.23  & 5.925   & 0.878   & 4.970   & Individual\,SPW    \\
338.39  & 6.705   & 1.001   & 5.877   & All\,SPW    \\
\hline
\end{tabular}
\end{table}
\footnotetext[1]{All images have been convolved to a common beam-size $4\farcs1\times2\farcs9$.}
\footnotetext[1]{Flux density determined using a common aperture of size $6{''}\times4{''}$ for all wavelengths.}
\clearpage

\end{document}